\documentclass[a4paper,11pt]{article}
\pdfoutput=1 

\usepackage{jcappub} 

\usepackage[T1]{fontenc} 
\usepackage{verbatim}
\usepackage[utf8]{inputenc}
\usepackage[american]{babel}
\usepackage{graphicx}
\usepackage{epsfig}
\usepackage{adjustbox}
\usepackage{booktabs}
\usepackage{multirow}
\usepackage{dcolumn}
\usepackage{amsmath}
\usepackage{amsfonts}
\usepackage{mathtools}
\usepackage{amssymb}
\usepackage{empheq}
\usepackage{subcaption} 
\usepackage{caption}
\usepackage{epstopdf}
\usepackage{bm}
\usepackage{bbm}
\usepackage{dsfont}
\usepackage{mathrsfs}
\usepackage{version}
\usepackage{enumerate}
\usepackage{braket}
\usepackage{enumitem}
\usepackage{transparent}
\usepackage{pifont}
\usepackage{colortbl}
\usepackage{rotating}
\usepackage{adjustbox}
\usepackage{cancel}
\usepackage{tabularx}
\usepackage{soul}
\usepackage{pdfpages}
\usepackage{microtype}
\usepackage{ulem}
\usepackage[table]{xcolor}
\usepackage{siunitx}
\usepackage{float} 
\usepackage[nocompress]{cite} 
\usepackage{marginnote}
\usepackage{relsize}
\usepackage{feynmp-auto}

\usepackage{hyperref}
\hypersetup{colorlinks=true,
linkcolor=navyblue,citecolor=coralred,urlcolor=green(munsell),pdfencoding=auto}

\definecolor{navyblue}{rgb}{0.0, 0.0, 0.5}
\definecolor{bleudefrance}{rgb}{0.19, 0.55, 0.91}
\definecolor{coralred}{rgb}{1.0, 0.25, 0.25}
\definecolor{royalblue}{rgb}{0.25, 0.41, 0.88}
\definecolor{cadmiumgreen}{rgb}{0.0, 0.42, 0.24}
\definecolor{green(munsell)}{rgb}{0.0, 0.66, 0.47}
\definecolor{blue-violet}{rgb}{0.54, 0.17, 0.89}
\definecolor{darkviolet}{rgb}{0.58, 0.0, 0.83}
\definecolor{orange(colorwheel)}{rgb}{1.0, 0.5, 0.0}
\definecolor{internationalorange}{rgb}{1.0, 0.31, 0.0}

\definecolor{magenta(process)}{rgb}{1.0, 0.0, 0.56}

\definecolor{darkspringgreen}{rgb}{0.09, 0.45, 0.27}

\definecolor{royalblue(web)}{rgb}{0.25, 0.41, 0.88}

\definecolor{cadmiumorange}{rgb}{0.93, 0.53, 0.18}

\definecolor{heliotrope}{rgb}{0.87, 0.45, 1.0}

\makeatletter
\renewcommand*{\@textcolor}[3]{%
\protect\leavevmode
\begingroup
\color#1{#2}#3%
\endgroup
}
\makeatother

\newcommand{\myfloatalign}{\centering}

\newcommand{\hp}{\hphantom}

\makeatletter
\newlength{\apb@width}
\newcommand{\autoparbox}[2][c]{\settowidth{\apb@width}{#2}\parbox[#1]{\apb@width}{#2}}

\makeatother

\DeclareCaptionLabelSeparator{colquad}{.\,\,\,}
\DeclarePairedDelimiter{\abs}{\lvert}{\rvert}

\makeatletter
\let\save@mathaccent\mathaccent
\newcommand*\if@single[3]{%
\setbox0\hbox{${\mathaccent"0362{#1}}^H$}%
\setbox2\hbox{${\mathaccent"0362{\kern0pt#1}}^H$}%
\ifdim\ht0=\ht2 #3\else #2\fi
}
\newcommand*\rel@kern[1]{\kern#1\dimexpr\macc@kerna}
\newcommand*\widebar[1]{\@ifnextchar^{{\wide@bar{#1}{0}}}{\wide@bar{#1}{1}}}
\newcommand*\wide@bar[2]{\if@single{#1}{\wide@bar@{#1}{#2}{1}}{\wide@bar@{#1}{#2}{2}}}
\newcommand*\wide@bar@[3]{%
\begingroup
\def\mathaccent##1##2{%
\let\mathaccent\save@mathaccent
\if#32 \let\macc@nucleus\first@char \fi
\setbox\z@\hbox{$\macc@style{\macc@nucleus}_{}$}%
\setbox\tw@\hbox{$\macc@style{\macc@nucleus}{}_{}$}%
\dimen@\wd\tw@
\advance\dimen@-\wd\z@
\divide\dimen@ 3
\@tempdima\wd\tw@
\advance\@tempdima-\scriptspace
\divide\@tempdima 10
\advance\dimen@-\@tempdima
\ifdim\dimen@>\z@ \dimen@0pt\fi
\rel@kern{0.6}\kern-\dimen@
\if#31
\overline{\rel@kern{-0.6}\kern\dimen@\macc@nucleus\rel@kern{0.4}\kern\dimen@}%
\advance\dimen@0.4\dimexpr\macc@kerna
\let\final@kern#2%
\ifdim\dimen@<\z@ \let\final@kern1\fi
\if\final@kern1 \kern-\dimen@\fi
\else
\overline{\rel@kern{-0.6}\kern\dimen@#1}%
\fi
}%
\macc@depth\@ne
\let\math@bgroup\@empty \let\math@egroup\macc@set@skewchar
\mathsurround\z@ \frozen@everymath{\mathgroup\macc@group\relax}%
\macc@set@skewchar\relax
\let\mathaccentV\macc@nested@a
\if#31
\macc@nested@a\relax111{#1}%
\else
\def\gobble@till@marker##1\endmarker{}%
\futurelet\first@char\gobble@till@marker#1\endmarker
\ifcat\noexpand\first@char A\else
\def\first@char{}%
\fi
\macc@nested@a\relax111{\first@char}%
\fi
\endgroup
}
\makeatother

\newcommand\ee{\end{equation}}
\newcommand\be{\begin{equation}}
\newcommand\eea{\end{eqnarray}}
\newcommand\bea{\begin{eqnarray}}
\newcommand{\bsp}{\begin{split}}
\newcommand{\esp}{\end{split}}
\newcommand{\bit}{\begin{itemize}[leftmargin=*]}
\newcommand{\eit}{\end{itemize}}
\newcommand{\ben}{\begin{enumerate}[leftmargin=*]}
\newcommand{\een}{\end{enumerate}}

\renewcommand{\emph}{\textit}

\newcommand\eq[1]{Eq.~\eqref{eq:#1}}
\newcommand\eqsI[1]{Eqs.~\eqref{eq:#1}}
\newcommand{\eqsII}[2]{Eqs.~\eqref{eq:#1}, \eqref{eq:#2}}
\newcommand{\eqsIII}[3]{Eqs.~\eqref{eq:#1}, \eqref{eq:#2}, \eqref{eq:#3}}
\newcommand{\eqsIV}[4]{Eqs.~\eqref{eq:#1}, \eqref{eq:#2}, \eqref{eq:#3}, \eqref{eq:#4}}

\newcommand{\iu}{\mathrm{i}}
\newcommand{\eu}{\mathrm{e}}
\newcommand{\dif}{\mathrm{d}}

\newcommand{\Tr}{\mathrm{Tr}\,}
\renewcommand{\vec}{\bm} 
\newcommand\vers[1]{\hat{\vec{#1}}}

\newcommand\eps{\varepsilon}

\def\gmgm{\gamma\gamma}
\def\zz{\zeta\zeta}
\def\ggg{\gamma\gamma\gamma}
\def\ggz{\zeta\gamma\gamma}
\def\gzz{\gamma\zeta\zeta}
\def\zzz{\zeta\zeta\zeta}
\def\mpl{M_{\rm P}}

\newcommand{\fnl}[1]{f_\textnormal{\textsc{nl}}^{#1}}

\newcommand{\threeG}[3]{{^{(3)}\Gamma^{#1}_{#2#3}}}

\def\<{\left\langle}
\def\>{\right\rangle}

\def\comment#1{}

\title{Graviton non-Gaussianities and Parity Violation in the EFT of Inflation}

\author[a]{Lorenzo Bordin,}
\author[b]{Giovanni Cabass}

\affiliation[a]{School of Physics \& Astronomy, University of Nottingham, University Park, Nottingham, NG7 2RD, UK}
\affiliation[b]{Max-Planck-Institut f\"{u}r Astrophysik, 
Karl-Schwarzschild-Str. 1, 85741 Garching, Germany}

\emailAdd{lorenzo.bordin@nottingham.ac.uk} 
\emailAdd{gcabass@mpa-garching.mpg.de}

\abstract{\noindent We study graviton non-Gaussianities in the EFT of Inflation. 
At leading (second) order in derivatives, the graviton bispectrum is fixed by Einstein gravity. 
There are only two contributions at third order. One of them breaks parity. 
They come from operators that directly involve the foliation: we then expect sizable non-Gaussianities in three-point 
functions involving both gravitons and scalars. However, we show that at leading order in slow roll 
the parity-odd operator does not modify these mixed correlators. 
We then identify the operators that can affect the graviton bispectrum 
at fourth order in derivatives. There are two operators that preserve parity. 
We show that one gives a scalar-tensor-tensor three-point function larger than the one computed in 
Maldacena, 2003 \cite{Maldacena:2002vr} if $M^2_{\rm P}A_{\rm s}/\Lambda^2\gg 1$ (where $\Lambda$ is the scale suppressing this 
operator and $A_{\rm s}$ the amplitude of the scalar power spectrum). 
There are only two parity-odd operators at this order in derivatives.}

\begin{document}
\maketitle
\flushbottom



\section{Introduction and summary of main results} 
\label{sec:introduction}

\noindent Upcoming CMB experiments will target the primordial tensor-to-scalar ratio $r$ to a sensitivity 
of $\sigma_r\sim 10^{-3}$ \cite{Matsumura:2013aja,Abazajian:2016yjj,Finelli:2016cyd,Delabrouille:2017rct,
Suzuki:2018cuy,Ade:2018sbj,Hanany:2019lle,Shandera:2019ufi,Abazajian:2019eic}. If a detection of vacuum 
fluctuations of gravity is achieved, the way towards constraints on tensor non-Gaussianities, 
and then on the \emph{interactions} of the graviton, will open. 
In contrast with the scalar sector, the situation for tensor perturbations is very constrained 
and, unless one considers exotic models with higher-spin degrees of freedom 
(see e.g.~\cite{Arkani-Hamed:2015bza,Lee:2016vti,Agrawal:2017awz,Bordin:2018pca,Goon:2018fyu,Dimastrogiovanni:2018gkl}), 
it is hard to go beyond Maldacena's result \cite{Maldacena:2002vr}. For instance, 
in \cite{Maldacena:2011nz} it was shown that, at leading order in slow roll, 
conformal invariance ensures that there are only two possible additional shapes 
beyond that first calculated in \cite{Maldacena:2002vr}, with only one actually appearing in the graviton bispectrum. 

The result can be circumvented if explicit couplings of the metric with the foliation are allowed. 
However, Ref.~\cite{Creminelli:2014wna} showed that, even if such couplings are turned on, 
at leading (second) order in derivatives the graviton power spectrum and bispectrum are fixed 
to be those calculated in \cite{Maldacena:2002vr}. 
It is therefore interesting to extend this analysis beyond the leading order in derivatives. 
We will do this by means of the Effective Field Theory of Inflation (EFTI) \cite{Cheung:2007st}, 
in which one can study in full generality the couplings of gravity with the fluctuations of the 
physical clock describing the slicing of spacetime during inflation. 

Let us summarize our approach and our results. We look at an expansion in 
powers of $H/\Lambda$ around Einstein gravity, where the Hubble parameter $H$ is the typical energy scale of tensor fluctuations 
during inflation and $\Lambda$ is some energy scale controlling the size of higher-derivative operators. 
First, we focus on the contributions to the graviton two- and three-point functions $\braket{\gmgm}$ and $\braket{\ggg}$ 
at next-to-leading order (NLO): these come from operators that carry three derivatives acting on the metric in unitary gauge. 
We identify two parity-odd operators that, starting quadratic in perturbations around an FLRW background, 
can correct both the tensor power spectrum and the tensor bispectrum. 
Only one gives a nonzero result on super-horizon scales. 
Moreover, there is an additional parity-even correction to the tensor bispectrum 
coming from an operator that starts cubic in perturbations. 

We then focus on mixed correlators from the two parity-odd quadratic operators. We show that, 
at leading order in the slow-roll parameter $\smash{\varepsilon = {-\dot{H}}/H^2}$, 
neither of them modifies the scalar-tensor-tensor and the tensor-scalar-scalar bispectra. 

We conclude with a discussion on the operators at next-to-next-to-leading order (NNLO) in derivatives. 
We show that only two parity-even operators (one starting quadratic and one cubic in perturbations) survive after 
integration by parts and field redefinitions. They involve direct couplings with the foliation. 
We compute the $\braket{\ggz}$ and $\braket{\gzz}$ three-point functions 
from the quadratic operator. It turns out that the latter is vanishing at late times, while the former is 
larger than the one computed in \cite{Maldacena:2002vr} if the ratio $H^2/\Lambda^2$ between the 
Hubble rate and the scale suppressing this operator is larger than 
$\varepsilon$ (equivalently, if $M^2_{\rm P}A_{\rm s}/\Lambda^2 \gg 1$, being $A_{\rm s}$ the amplitude of the scalar power spectrum). 
Finally, we show that there are two parity-odd operators starting cubic in perturbations. 

The paper is organized as follows. 
In Section~\ref{sec:redundant_operators} we briefly review the results of \cite{Creminelli:2014wna,Bordin:2017hal} and 
discuss the EFTI predictions for the correlators involving the graviton at leading order in derivatives. 
In Section~\ref{sec:four_operators} we look in more detail at the operators that enter at NLO in derivatives. 
We present the full results for the correlation functions in Section~\ref{sec:NGs}. 
In Section~\ref{sec:NNLO} we discuss the operators at fourth order in derivatives. 
In Section~\ref{sec:CRs} we check the soft limits for the correlators of Sections~\ref{sec:NGs} 
and~\ref{sec:NNLO}, and we conclude in Section~\ref{sec:conclusions}. 
In Appendix~\ref{app:notation} we summarize our notation and conventions. 
Appendices~\ref{app:appendix-minus_1}, \ref{app:cubic_action_3DCS}, 
\ref{app:appendix-0} and \ref{app:appendix-C} contain some details of the 
calculations carried out in the three main sections.

\section{Redundant operators at leading order in derivatives} 
\label{sec:redundant_operators}

\subsection{Field redefinitions and graviton bispectrum}
\label{subsec:field_redefinitions_and_graviton_bispectrum}

\noindent To simplify the tensor sector as much as possible, one can perform field redefinitions that decay at late times. 
At leading order in derivatives the graviton action can be put in the 
Einstein-Hilbert form, and consequently both the graviton power spectrum and bispectrum are completely fixed 
by Maldacena's result \cite{Maldacena:2002vr,Creminelli:2014wna}. 

To go beyond this we will consider interactions with more than two derivatives acting on the metric. 
Stopping at fourth order in derivatives we can write our action as 
\begin{equation}
\label{eq:summary_redundancies-A}
S = S_0 + \sum_I S_{\Lambda_I} + \sum_I S_{\widebar{\Lambda}^2_I}\,\,,
\end{equation}
where 
\begin{subequations}
\label{eq:summary_redundancies-B}
\begin{align}
S_0 &= \frac{M^2_{\rm P}}{2}\int\dif^4x\,\sqrt{-g}\,\bigg(R-\frac{2\dot{H}}{N^2} 
- 2(3H^2+\dot{H})\bigg)\,\,, \label{eq:summary_redundancies-B-1} \\
S_{\Lambda_I} &= {M^2_{\rm P}}\int\dif^4x\,\sqrt{-g}\,\frac{{\cal O}_{1,I}}{\Lambda_{I}}\,\,, \label{eq:summary_redundancies-B-2} \\ 
S_{\widebar{\Lambda}^2_I} &= {M^2_{\rm P}}\int\dif^4x\,\sqrt{-g}\,\frac{{\cal O}_{2,I}}{\widebar{\Lambda}^2_{I}}\,\,. 
\label{eq:summary_redundancies-B-3}
\end{align}
\end{subequations} 
The operators $\smash{{\cal O}_{1,I}}$ and $\smash{{\cal O}_{2,I}}$ 
are constructed by combining the perturbation of the lapse function 
$\delta\!N=N-1$, of the extrinsic curvature $\delta\!K_{\mu\nu} = K_{\mu\nu} - Hh_{\mu\nu}$ and of its trace $\delta\!K = K-3H$ 
(with $h^\mu_{\hp{\mu}\nu}=\delta^\mu_{\hp{\mu}\nu}+n^\mu n_\nu$ being the projector on the hypersurfaces of 
constant time and $n^\mu = (1,-N^i)/N$, $n_\mu = -N\delta_\mu^0$ their normal vector), 
the $3$-dimensional Riemann tensor $\smash{{^{(3)}}\!R_{\rho\sigma\mu\nu}}$, the ADM ``acceleration'' vector 
$A^\mu = n^\nu\nabla_\nu n^\mu = h^{\mu\nu}\nabla_\nu\log N$, 
and the derivative of the lapse projected along the normal to the foliation $\smash{V=n^\mu\nabla_\mu N}$. 

The coefficients $\smash{\Lambda_I}$ and $\smash{\widebar{\Lambda}^2_I}$ are in general time-dependent. 
In this paper we assume that their variation in time is small, suppressed by 
$\eps$ (we refer to \cite{Finelli:2018upr} for a more detailed discussion about their time dependence). 
Consistently with this, we will assume an exact de Sitter background. 

The rest of the paper studies what operators $\smash{{\cal O}_{1,I}}$ and $\smash{{\cal O}_{2,I}}$ affect 
the graviton bispectrum. As we explained in the introduction, all the operators that we will discuss 
involve direct couplings between the metric and the foliation. 
One might wonder if these couplings are slow-roll-suppressed. We can easily 
see that this is not the case by looking, for example, at a simple $P(X,\phi)$ theory. In operators of the form 
$X^n$ many of the legs can be evaluated on the background. This will give additional factors of $\dot{\phi}^{1/2}$ 
that are large with respect to $H$, i.e.~the size of derivatives acting on field fluctuations around horizon crossing 
($\smash{H/{\dot{\phi}}^{1/2} = {\cal O}(10^{-2})}$ from the normalization of the scalar power spectrum). 
For a more detailed discussion, see e.g.~\cite{Creminelli:2003iq,Cheung:2007st,Baumann:2011su}.

\subsection{Mixed tensor-scalar correlators}
\label{subsec:mixed_tensor_scalar_correlators}

\noindent When discussing graviton non-Gaussianities it is also interesting to look 
at three-point functions involving both tensor and scalar modes. 
However, as it has been discussed in \cite{Bordin:2017hal}, 
these are not very constrained already at the two-derivative level, 
so it is difficult to make general statements for them. 
A simplification, however, occurs if we consider only those contributions that the nonlinear realization of 
time diffeomorphisms links to the modification of the graviton power spectrum. 
In the following we argue that these are the first one should constrain in case of a detection of the tensor power spectrum. 
\begin{itemize}[leftmargin=*]
\item Let us assume that a difference in the power spectra of left- and right-handed 
graviton helicities is detected, for example via a nonzero correlation 
of CMB $E$- and $B$-modes. We are then guaranteed that a parity-odd operator is present in the action $S$ of 
\eqsII{summary_redundancies-A}{summary_redundancies-B}, 
and we can go to look for a signal in observables like the $EBB$ or $BEE$ bispectra 
evaluated at configurations that would vanish if parity is conserved 
(see also \cite{Biagetti:2020lpx} for a recent study of signatures in galaxy intrinsic alignments). 
\item Let us then consider the parity-conserving scenario. The fact that at leading order in derivatives the tensor 
power spectrum is univocally fixed by the Hubble rate leads to a ``consistency relation'' between the tensor tilt 
and the tensor-to-scalar ratio, $r=-8n_{\rm t}$. At higher orders, instead, the amplitude of $\braket{\gmgm}$ 
depends also on other EFT coefficients: the ``consistency relation'' will be broken by an amount controlled by 
$H^2/\Lambda^2$ (as we will see, e.g.,~in Section~\ref{subsec:NNLO_parity_even}). Hence, a detection 
of $r\neq-8n_{\rm t}$ implies the presence of higher-derivatives operators. We should therefore look for their 
imprint in observables that probe $\braket{\ggz}$ and $\braket{\gzz}$: these will be crucial to determine $\eps$ since they 
give a handle on the combination $H^2/\Lambda^2$. 
\end{itemize}

\section{Operators involving the graviton at NLO}
\label{sec:four_operators}

\subsection{Parity-even operators}
\label{subsec:parity_even_operators_NLO}

\noindent Let us first focus on operators that do not break parity. 
It is not possible to write down a parity-even correction to the tensor power spectrum at third order in 
derivatives \cite{Creminelli:2014wna,Cannone:2014uqa,Bordin:2017hal}: 
at this order we have only the cubic operator $\delta\!K^{\alpha\gamma} 
\delta\!K_\gamma^{\hp{\gamma}\rho}\delta\!K_{\rho\alpha} \supset\dot{\gamma}_{ij} 
\dot{\gamma}_{jk}\dot{\gamma}_{ki}/8$ \cite{Bordin:2017hal}, i.e.~ 
\begin{equation}
\label{eq:deltaK_ij_cubed}
S_{\Lambda_1} = M^2_{\rm P}\int\dif^4x\,\sqrt{-g}\,\frac{\delta\!K_\mu^{\hp{\mu}\nu}
\delta\!K_{\nu}^{\hp{\nu}\rho} \delta\!K_\rho^{\hp{\rho}\mu}}{\Lambda_1}\,\,. 
\end{equation}
This operator will contribute to the graviton bispectrum. 

Notice that, na{\" i}vely, there is an additional operator at this order in derivatives. 
Consider the Gauss-Bonnet term $R^2 - 4R^{\sigma\nu}R_{\sigma\nu}+R^{\rho\sigma\mu\nu}R_{\rho\sigma\mu\nu}$: 
in four dimensions, it is equal to the four-divergence of a current, i.e.~ 
\begin{equation}
\label{eq:GB-1}
{{\cal G}\!{\cal B}}\equiv R^2 - 4R^{\sigma\nu}R_{\sigma\nu}+R^{\rho\sigma\mu\nu}R_{\rho\sigma\mu\nu} = \nabla_\mu G^\mu\,\,.
\end{equation}
This current $G^\mu$ is not a four-vector. However, the fact that the Gauss Bonnet combination is independent 
of the chosen coordinate system tells us that $G^\mu$ transforms as such under $\smash{\int\dif^4x\,\sqrt{-g}}$. 
More precisely, $\smash{\int\dif^4x\,\sqrt{-g}}\,G^0$ is invariant under spatial diffeomorphisms and 
transforms as the zeroth component of a four-vector under time diffeomorphisms. It is therefore a 
legit object to be added to the EFTI action. 

A manageable expression for $G^\mu$ in terms of metric components and Christoffel symbols 
is not easy to derive \cite{Yale:2010jy}. This obscures a bit the fact that, indeed, $G^0$ is build from 
the usual geometric objects employed in the EFTI. 
Let us show it. We focus on gravitons only,\footnote{We do this only to simplify the calculations: the same conclusion holds 
if we include also scalars.} so that the relation $a^3\nabla_\mu G^\mu = \partial_\mu(a^3 G^\mu)$ holds. 
At quadratic order in $\gamma_{ij}$, and assuming a constant Hubble parameter for simplicity, 
the Gauss-Bonnet term contains the following structures: 
\begin{equation}
\label{eq:GB-5}
\begin{split}
a^3{{\cal G}\!{\cal B}} \supset &\,\big({-a^3}H\dot{\gamma}_{ij}
\dot{\gamma}_{ij}\big)^\cdot\,, 
\,\,\big({-aH}\partial_k\gamma_{ij}\partial_k\gamma_{ij}\big)^\cdot\,, 
\,\,\big(2a\dot{\gamma}_{ij}\partial^2\gamma_{ij}\big)^\cdot\,, 
\,\,\partial_i \big(a^3 G^i\big)\,\,. 
\end{split}
\end{equation}
In the first combination we recognize the four-divergence 
$\smash{\nabla_\mu({-4H}\delta\!K_{\rho\sigma} \delta\!K^{\rho\sigma} n^\mu)}$. 
Using the perturbative expressions for $^{(3)}\!R$ and $^{(3)}\!R_{ij}$ 
we recognize $\smash{\nabla_\mu(4H{{^{(3)}}\!R}n^\mu)}$ and 
$\smash{\nabla_\mu(8\,\delta\!K_{\rho\sigma}{{^{(3)}}\!R^{\rho\sigma}}n^\mu)}$ 
in the second and third combinations, respectively. The first two are lower-derivative operators: 
one can deal with them with the field redefinitions discussed in 
Section~\ref{subsec:field_redefinitions_and_graviton_bispectrum}. The third one can be integrated 
by parts to give lower-derivative operators and operators involving scalar modes 
(see e.g.~\eq{five_operators-B} of Appendix~\ref{app:appendix-0}). 

At cubic order in fluctuations we would find that $G^0$ contains also a term of the form 
$\delta\!K^{\alpha\gamma}\delta\!K_\gamma^{\hp{\gamma}\rho}\delta\!K_{\rho\alpha} n^\mu$. 
In summary this tells us that there is no need to consider this operator in the EFTI action.

\subsection{Parity-odd operators}
\label{subsec:parity_odd_operators_NLO}

\noindent We can now focus on parity-odd operators. At third order in derivatives there are three operators that, 
in principle, contribute to $\braket{\gmgm}$. These are\footnote{The signs and overall numerical 
factors in \eqsI{three_quadratic_parity_odd} are chosen to reproduce the actions used in 
\cite{Creminelli:2014wna,Bartolo:2017szm}. More precisely, 
\eqsII{three_quadratic_parity_odd-1}{three_quadratic_parity_odd-2} reproduce Eq.~(21) of \cite{Creminelli:2014wna} upon 
identification of their $\alpha/\Lambda$, $\beta/\Lambda$ with $1/\Lambda_2$, $1/\Lambda_3$, 
and for $\Lambda_4 = M_{\rm CS}$ we match the action used in \cite{Bartolo:2017szm} at leading order in slow roll.} 
\begin{subequations}
\label{eq:three_quadratic_parity_odd}
\begin{align}
S_{\Lambda_2} &= \frac{M^2_{\rm P}}{2}\int{\rm d}^4x\,\sqrt{-g}\,\frac{1}{\Lambda_2}\,
\frac{\vec{e}^{\mu\nu\rho\sigma}n_\mu}{N}\,D_\nu\delta\!K_{\rho\lambda}\delta\!K^\lambda_{\hp{\lambda}\sigma} 
\,\,, \label{eq:three_quadratic_parity_odd-1} \\
S_{\Lambda_3} &= {\frac{M^2_{\rm P}}{2}}\int{\rm d}^4x\,\sqrt{-g}\,\frac{1}{\Lambda_3}\,
{\vec{e}^{0 ijk}}\,\Bigg(\frac{\threeG{l}{i}{m}\partial_j\threeG{m}{k}{l}}{2} 
+ \frac{\threeG{l}{i}{m}\threeG{m}{j}{n}\threeG{n}{k}{l}}{3}\Bigg)\,\,, \label{eq:three_quadratic_parity_odd-2} \\
S_{\Lambda_4} &= {M^2_{\rm P}}\int\dif^4x\,\sqrt{-g}\,
\frac{1}{\Lambda_4}\,\vec{e}^{0\alpha\beta\gamma}\Bigg(\frac{\Gamma^\sigma_{\alpha\nu}\partial_\beta\Gamma^\nu_{\gamma\sigma}}{2} 
+ \frac{\Gamma^\sigma_{\alpha\nu}\Gamma^\nu_{\beta\lambda}\Gamma^\lambda_{\gamma\sigma}}{3}\Bigg)
\,\,. \label{eq:three_quadratic_parity_odd-3}
\end{align}
\end{subequations} 
The volume form $\smash{\vec{e}_{\mu\nu\rho\sigma}}$ ($\smash{\vec{e}^{\mu\nu\rho\sigma}}$) is equal to 
$\smash{\sqrt{-g}\,\epsilon_{\mu\nu\rho\sigma}}$ ($\smash{{-\epsilon_{\mu\nu\rho\sigma}}/\sqrt{-g}}$), 
where $\smash{\epsilon_{\mu\nu\rho\sigma}}$ is the Levi-Civita symbol such that $\epsilon_{0ijk} = \epsilon_{ijk}$, $\epsilon_{0123}=1$. 
The operator in \eq{three_quadratic_parity_odd-2} is the Chern-Simons term for the metric on the hypersurfaces of 
constant time.\footnote{Ref.~\cite{Takahashi:2009wc} discusses this term in 
the context of single-field inflation and Ho{\v r}ava-Lifshitz gravity. Notice however that the term itself is not present in 
the action: the parity-odd operator considered is $\smash{\vec{e}^{\mu\nu\rho\sigma}n_\mu 
{{^{(3)}}\!R_{\nu}^{\hp{\nu}\lambda}}D_\rho{{^{(3)}}\!R_{\lambda\sigma}}}$, 
which is of subleading in derivatives with respect to \eq{three_quadratic_parity_odd-2}.} 
The operator in \eq{three_quadratic_parity_odd-3}, instead, is the zeroth component of the current $K^\mu$ that satisfies 
\begin{equation}
\label{eq:chern_simons_current_definition-1}
\begin{split}
\nabla_\mu K^\mu = \frac{1}{4}\vec{e}^{\mu\nu\alpha\beta}R^\sigma_{\hp{\sigma}\rho\alpha\beta}R^\rho_{\hp{\rho}\sigma\mu\nu}\,\,, 
\end{split}
\end{equation}
where
\begin{equation}
\label{eq:chern_simons_current_definition-2}
K^\mu = 2\vec{e}^{\mu\alpha\beta\gamma}\Bigg(\frac{\Gamma^\sigma_{\alpha\nu}\partial_\beta\Gamma^\nu_{\gamma\sigma}}{2} 
+ \frac{\Gamma^\sigma_{\alpha\nu}\Gamma^\nu_{\beta\lambda}\Gamma^\lambda_{\gamma\sigma}}{3}\Bigg)\,\,.
\end{equation} 
The contributions to the tensor power spectrum of the two operators of 
\eqsII{three_quadratic_parity_odd-1}{three_quadratic_parity_odd-2}, and the contribution to 
the scalar-tensor-tensor three-point function of the third one, were computed respectively in \cite{Creminelli:2014wna} 
and \cite{Satoh:2010ep,Bartolo:2017szm} (see also \cite{Cordova:2017zej} 
for a discussion about the contribution to $\braket{\ggz}$ of $S_{\Lambda_4}$). 

Notice that, similarly to $G^\mu$, $K^\mu$ is not a four-vector. However one can again show that $K^0$ is 
a legit operator of the EFTI: its transformation properties under $\int\dif^4x\,\sqrt{-g}$ 
are the same as those of the zeroth component of a four-vector. 
Since $\int\dif^4x\,\sqrt{-g}\,K^0$ is invariant under spatial diffeomorphisms one might wonder if 
it can be decomposed in more ``fundamental'' building blocks via the $3+1$ splitting of spacetime, as we did with $G^0$. 
This is indeed the case: in Appendix~\ref{app:appendix-minus_1} we show that, 
when we consider tensor modes only, it is always possible to write $S_{\Lambda_4}$ as a combination of $S_{\Lambda_2}$ 
and $S_{\Lambda_3}$ (including scalar modes would require additional operators to fully decompose $S_{\Lambda_4}$, 
but we are not interested in them in this paper). For this reason we will not consider $S_{\Lambda_4}$ in the following. 

Before proceeding, let us discuss whether it is possible to remove these operators via field redefinitions. 
For simplicity we work perturbatively in $1/\Lambda_I$. 
Since these operators are third-order in derivatives, and the variation of $S_0$ 
carries at least two derivatives due to the variation of the Einstein-Hilbert action, 
we are forced to consider field redefinitions that carry at most one derivative. 
Moreover, these field redefinitions must include the volume form $\vec{e}^{\mu\nu\rho\sigma}$ since $S_0$ is parity-conserving. 
It is then straightforward to realize that we cannot write down any redefinition 
$g^{\mu\nu}\to g^{\mu\nu}+\delta g^{\mu\nu}$ that would keep the metric a symmetric tensor. 

We also notice that the we cannot use field redefinitions that keep the Einstein-Hilbert action unchanged. 
These are redefinitions of the form $g^{\mu\nu}\to g^{\mu\nu}+2\nabla^{(\mu}\xi^{\nu)}$ where, due to the 
invariance of the EFTI action under spatial diffeomorphisms, it is enough to consider 
$\xi^\mu = F n^\mu$ \cite{Bordin:2017hal}. Given that the Einstein-Hilbert action does not vary 
(thanks to the Bianchi identity), we are allowed to take $F$ to be at most of second order in derivatives. 
Even so, we cannot construct, from the building blocks of the EFTI, 
a function $F$ that contains $\vec{e}^{\mu\nu\rho\sigma}$ while remaining at second order in derivatives. 

We conclude this section by checking if there are parity-odd operators at cubic order 
in perturbations and third order in derivatives that contribute to $\braket{\ggg}$. 
It is straightforward to convince ourselves that there is no such operator. 
Let us work in term of the graviton fluctuations. The presence 
of $\epsilon_{ijk}\sim\vec{e}^{\mu\nu\rho\sigma}n_\mu$ allows us to consider also 
$\partial_k\dot{\gamma}_{ij}\sim D_\mu\delta\!K_{\rho\sigma}$ in addition to 
$\partial^2\gamma_{ij}\sim{{^{(3)}}\!R_{\mu\nu}}$ and $\dot{\gamma}_{ij}\sim\delta\!K_{\mu\nu}$ 
We also want three powers of $\gamma_{ij}$ to contract with $\epsilon_{ijk}$. However 
there are too many indices that would remain free unless we add additional spatial derivatives or 
scalar modes through the ADM acceleration vector $A^\mu$ (for the same reason, we still cannot consider 
$\partial_k\partial_l\gamma_{ij}\sim{{^{(3)}}\!R_{\rho\sigma\mu\nu}}$ as a building block).

\section{Graviton non-Gaussianities at NLO}
\label{sec:NGs}

\noindent In this section we compute the super-horizon correlation functions via the in-in formalism. 
Before doing that, however, it is worth to estimate the size of these new 
contributions with respect to the non-Gaussianities coming from the minimal action $S_0$. 
This is especially important for the mixed scalar-tensor-tensor and tensor-scalar-scalar 
three-point functions. Indeed we expect that they are slow-roll-enhanced with respect to 
those of \cite{Maldacena:2002vr}. In Section~\ref{subsec:in_in} we will see that, 
at variance with this expectation, the mixed three-point functions 
from $S_{\Lambda_I}$, $I=2,3$, vanish at leading order in slow-roll parameters.

\subsection{Estimates}
\label{subsec:estimates}

\noindent Let us first review the non-Gaussianities from the minimal action $S_0$. 
It is simpler to work in flat gauge: let us reintroduce the Stueckelberg field $\pi$ via $t\to t+\pi$ 
(see Appendix~\ref{app:notation} for details). 
At leading order in slow roll, on super-horizon scales $\zeta$ and $\pi$ are related by $\zeta = -H\pi$. 
After reintroducing $\pi$ we can solve for the constraints: they are $\delta\!N = \varepsilon H\pi$, 
$a^{-2}\partial^2N_{\rm L} = -\varepsilon H \dot{\pi}$ \cite{Maldacena:2002vr,Cheung:2007sv}. 
At this point we can estimate the size of non-Gaussianities by comparing the cubic vertices to the 
quadratic Lagrangians for $\pi$ and for $\gamma_{ij}$. The quadratic action for the Stueckelberg field is of order $\varepsilon$: 
schematically, $S_0|_{\pi\pi}\sim \varepsilon\pi\pi$. 
The quadratic action for $\gamma_{ij}$ is instead $S_0|_{\gmgm}\sim\gmgm$. What about the cubic vertices? 
\begin{itemize}[leftmargin=*]
\item We first focus on the three-graviton vertex. It comes from the 
three-Ricci scalar ${^{(3)}}\!R$ \cite{Maldacena:2002vr,Maldacena:2011nz}. 
Hence we have $S_0|_{\ggg}\sim\ggg$. Since $S_0|_{\gmgm}\sim\gmgm$, we find 
$\braket{\ggg}_0\sim\braket{\gmgm}_0\braket{\gmgm}_0$. 
\item At leading order in slow roll the cubic vertex with one scalar and two gravitons comes from plugging the 
constraints in the Einstein-Hilbert action (which, being invariant under all 
diffeomorphisms, does not contain the $\pi$ field by itself). 
We then have at most $S_0|_{\pi\gamma\gamma}\sim\varepsilon\pi\gamma\gamma$: comparing with $S_0|_{\gmgm}\sim\gmgm$ 
we find that $\braket{\ggz}_0$ is of order $\varepsilon\braket{\gmgm}_0\braket{\zz}_0$. 
\item The cubic vertex with two scalars and one graviton comes from the term $-M^2_{\rm P}\dot{H}/N^2$ in $S_0$, 
which contains $\smash{g^{\mu\nu}\nabla_\mu\pi\nabla_\nu\pi\supset {-a^{-2}}\gamma_{ij}\partial_i\pi\partial_j\pi}$. 
From this we see that $S_0|_{\gamma\pi\pi}\sim\varepsilon\gamma\pi\pi$: 
since $S_0|_{\pi\pi}\sim\varepsilon\pi\pi$, we find $\braket{\gzz}_0 \sim \braket{\gmgm}_0\braket{\zz}_0$. 
\end{itemize}

We are now in the position to do the same estimates for $\smash{S_{\Lambda_I}}$. 
For $\smash{S_{\Lambda_1}}$ we care only about the graviton bispectrum. 
Since $\delta\!K^{\alpha\gamma}\delta\!K_\gamma^{\hp{\gamma}\rho}\delta\!K_{\rho\alpha}$ is simply 
$\dot{\gamma}_{ij}\dot{\gamma}_{jk}\dot{\gamma}_{ki}/8$ at cubic order in $\gamma_{ij}$, the estimate is very simple: 
we expect that $\braket{\ggg}_{\Lambda_1}/\braket{\ggg}_{0}$ is of order $H/\Lambda_1$. 
What about the parity-odd operators? Let us first consider their contribution to $\braket{\ggg}$. 
Also in this case we have that, for both operators, the cubic vertex with three gravitons carries 
an extra derivative, leading to a simple $H/\Lambda_I$ suppression. 

Things are a bit more tricky for correlators involving scalar modes. The reason is that we first need to study 
how the solution for the constraints is modified by the presence of the new operators (we emphasize that we work perturbatively 
in $1/\Lambda_I$ also in the solution of the constraint equations for $\delta\!N$ and $N_{\rm L}$). 
We will study the two operators separately, and summarize the results in Tab.~\ref{tab:estimates}.

\begin{table}
\myfloatalign
\caption[.]{Expected size of three-point functions for $S_{\Lambda_I}$, $I=2,3$, with respect to Maldacena's ones. 
We see that an enhancement by $1/\varepsilon$ is expected in both 
$\braket{\ggz}$ and $\braket{\gzz}$. As we discuss in Section~\ref{subsec:in_in}, 
the actual in-in calculation shows that such enhancement is absent.} 
\label{tab:estimates}
\centering
\medskip
\begin{tabular}{cccc}
\toprule
 & $\braket{\ggg}_{\Lambda}/\braket{\ggg}_{0}$ 
 & $\braket{\ggz}_{\Lambda}/\braket{\ggz}_{0}$ 
 & $\braket{\gzz}_{\Lambda}/\braket{\gzz}_{0}$ \\ 
\midrule
$S_{\Lambda_2}$	& $\dfrac{H}{\Lambda_2}$ & $\dfrac{H}{\Lambda_2\varepsilon}$ & $\dfrac{H}{\Lambda_2\varepsilon}$ \\ [2ex]
$S_{\Lambda_3}$	& $\dfrac{H}{\Lambda_3}$ & $\dfrac{H}{\Lambda_3\varepsilon}$ & $\dfrac{H}{\Lambda_3\varepsilon}$ \\ 
\bottomrule
\end{tabular}
\end{table}

\subsubsection*{$\vec{S_{\Lambda_2}}$}

\noindent The Stueckelberg trick for the operator of 
\eq{three_quadratic_parity_odd-1} is reviewed in Appendix~\ref{app:appendix-C}. 
All mixings of the $\pi$ field with the constraints vanish due to the presence of $\vec{e}^{\mu\nu\rho\sigma}n_\mu$. 
Moreover, it is straightforward to see that the modification of the quadratic action for $\delta\!N$ and $N_{\rm L}$ 
is not slow-roll-enhanced with respect to the contribution coming from $S_0$. 
Therefore the solution of the constraints in terms of $\pi$ is the same as Maldacena's one 
up to a change of order $H/\Lambda_2$ in their normalization. I.e.~we have, schematically, that 
\begin{equation}
\label{eq:constraints_schematic}
\delta\!N, N_{\rm L}\sim\varepsilon\pi\bigg(1+\frac{H}{\Lambda_2}\bigg)\,\,.
\end{equation}

Let us then focus on the cubic vertices containing both $\gamma_{ij}$ and $\pi$. 
Since $S_{\Lambda_2}$ is not invariant under time diffeomorphisms we expect these 
vertices to be present even before we plug in the solution for the constraints. 
This is shown explicitly in Appendix~\ref{app:appendix-C}. As expected, these vertices give (schematically) 
$S_{\Lambda_2}|_{\pi\gamma\gamma}\sim (H/\Lambda_2)\,\pi\gamma\gamma$, 
$S_{\Lambda_2}|_{\gamma\pi\pi}\sim (H/\Lambda_2)\,\gamma\pi\pi$. 
Hence, at leading order in slow roll and $H/\Lambda_2$ we can forget about 
$\delta\!N$ and $N_{\rm L}$, and just take $N=1$, $N^i=0$ throughout 
(we emphasize that, had there been a modification of the constraints 
at order $H/\Lambda_2$ that was not slow-roll-suppressed, we should have worried 
about having to substitute the constraints in $S_0$ as well, in order to capture all the relevant interactions). 

Then, taking the ratios $S_{\Lambda_2}|_{\pi\gamma\gamma}/S_0|_{\gamma\gamma}$ 
and $S_{\Lambda_2}|_{\gamma\pi\pi}/S_0|_{\pi\pi}$ we see that we should expect 
both $\braket{\ggz}_{\Lambda_2}/\braket{\ggz}_{0}$ and $\braket{\gzz}_{\Lambda_2}/\braket{\gzz}_{0}$ to be of 
order $(H/\Lambda_2)\,\varepsilon^{-1}$.

\subsubsection*{$\vec{S_{\Lambda_3}}$}

\noindent The three-dimensional Chern-Simons term is the simplest of the two operators. 
In this case we can work directly in $\zeta$ gauge: indeed, we see that in \eq{three_quadratic_parity_odd-2} 
the lapse and shift constraints do not appear. First, this tells us that the solution of 
$\delta\!N$ and $N_{\rm L}$ in terms of $\zeta$ is 
unchanged: after we plug them into $S_0$ to get the quadratic action for $\zeta$ we can forget about them. 
Using the same logic as we did for $S_{\Lambda_2}$, we then expect that the $\braket{\ggz}$ and $\braket{\gzz}$ 
three-point functions are suppressed by $H/\Lambda_3$ but enhanced by $1/\varepsilon$ with respect to those from $S_0$.

\subsection{Calculation via in-in formalism}
\label{subsec:in_in}

\noindent After these estimates, we are ready to look at the full results of the in-in calculation for these correlation functions. 

Before doing that let us briefly address the single-field consistency relations. 
Since the operators we consider are higher-derivative ones 
we are guaranteed that there is no modification of the squeezed correlation functions 
at order $1/q^3$ or $1/q^2$ ($\vec{q}$ being the long mode). The consistency relations are then 
satisfied if we take into account how the new operators modify the de Sitter modes for the 
$\pi$ and $\gamma$ fields once we compute the non-Gaussianities from $S_0$: we discuss this 
in more detail in Section~\ref{sec:CRs}.

\subsubsection*{$\vec{S_{\Lambda_1}}$}

\noindent We start from $S_{\Lambda_1}$. For the correction to the graviton bispectrum we find 
\begin{equation} 
\label{eq:S_Lambda_1} 
\Delta\!\braket{\gamma^{s_1}_{\vec{k}_1}\gamma^{s_2}_{\vec{k}_2}\gamma^{s_3}_{\vec{k}_3}}' = 
\frac{3}{8}{\frac{H}{\Lambda_1}}\frac{H^4}{\mpl^4} 
\frac{\epsilon_{ij}^{s_1}({-\vec{k}_1})\epsilon_{jk}^{s_2}({-\vec{k}_2})\epsilon_{ki}^{s_3}({-\vec{k}_3)}}{k_1 k_2 k_3\, k_T^3 }\,\,, 
\end{equation}
where $\smash{k_T\equiv\sum_i k_i}$ and $\smash{\epsilon_{ij}^{s}(\vec{k})}$ are the graviton polarization tensors 
(see Eq.~\eqref{eq:polarization_tensors_higher_spins} for the circularly-polarized tensors). 
Comparing with Eq.~(4.15) of \cite{Maldacena:2002vr} we see the expected $H/\Lambda_1$ suppression.

\subsubsection*{$\vec{S_{\Lambda_2}}$}

\noindent Let us move to $S_{\Lambda_2}$. The details of the calculation of the flat-gauge action 
are contained in Appendix~\ref{app:appendix-C}. The in-in calculation shows that 
$\braket{\gamma\gamma\gamma}$ vanishes super-horizon scales, as does $\braket{\gamma\gamma}$. 
We have also checked that at leading order in slow roll the corrections to $\braket{\ggz}$ and 
$\braket{\gzz}$ vanish: there is no $1/\varepsilon$ enhancement with respect to Maldacena's result. 

The only contribution of \eq{three_quadratic_parity_odd-1} is to add a field-dependent phase to 
the wavefunction of the universe on the boundary of de Sitter spacetime.\footnote{This can be seen 
by taking the imaginary part of \eq{in_in_master-B} after dropping the commutator. See \cite{Liu:2019fag} for a 
discussion of parity-violating signatures in the scalar sector, and how local interactions 
only give corrections to the phase of the wavefunction (and are then unobservable in correlators) 
at leading order in slow roll.} While the operator does not contribute to the vacuum expectation 
values of scalar and tensor fluctuations, it can therefore affect the scalar product between the vacuum and another state.

\subsubsection*{$\vec{S_{\Lambda_3}}$}

\noindent This operator gives a fractional correction to the power spectrum of 
the two helicities $\gamma^{s}_{\vec{k}}$ of the graviton proportional to $\lambda_s H/\Lambda_3$ 
($s=\pm$ and $\lambda_\pm = \pm1$) \cite{Satoh:2008ck,Satoh:2010ep,Wang:2012fi,Creminelli:2014wna,Bartolo:2017szm}. 

Interestingly there is also a nonzero modification of the graviton bispectrum 
(the parity-breaking operator $WWW^\ast$ considered in \cite{Maldacena:2011nz,Soda:2011am,Shiraishi:2011st} 
does not affect $\braket{\gamma\gamma\gamma}$ since it contributes to the wavefunction of the universe via a pure phase). 
We can write it as 
\begin{equation}
\label{eq:S_Lambda_3_in_in-A}
\Delta\!\braket{\gamma^{s_1}_{\vec{k}_1}\gamma^{s_2}_{\vec{k}_2}\gamma^{s_3}_{\vec{k}_3}}' = 
\frac{\pi}{16}\frac{H}{\Lambda_3}\frac{H^4}{\mpl^4} 
\frac{\epsilon^{s_1}_{il}({-\vec{k}_1})\epsilon^{s_2}_{jm}({-\vec{k}_2})\epsilon^{s_3}_{kn}({-\vec{k}_3}) 
T_{ijk}^{lmn}(\vec{k}_1,\vec{k}_2,\vec{k}_3)}{k_1^3k_2^3k_3^3}\,\,,
\end{equation} 
where $\smash{T_{ijk}^{lmn}(\vec{k}_1,\vec{k}_2,\vec{k}_3)}$ is symmetric under all 
exchanges of the form $\smash{(im\leftrightarrow jl,\vec{k}_1\leftrightarrow\vec{k}_2),\dots}$ 
and has dimension of a momentum to the third power. It can be constructed from the $\smash{6}$ 
permutations of the cubic vertex coming from $S_{\Lambda_3}$, which we write down in Appendix~\ref{app:cubic_action_3DCS}. 
This is however too generic. This expression does not make use of the fact that we are dealing with a parity-breaking operator, 
and that this operator is of cubic order in derivatives. In the case of the bispectrum from $\smash{S_0}$, we can write the 
``amplitude'' part as \cite{Maldacena:2002vr,Pajer:2020wxk} 
\begin{equation}
\label{eq:S_Lambda_3_in_in-B}
A_{S_0}(1^{s_1},2^{s_2},3^{s_3})\propto 
\epsilon^{s_1}_{il}({-\vec{k}_1})\epsilon^{s_2}_{jm}({-\vec{k}_2})\epsilon^{s_3}_{kn}({-\vec{k}_3})\, 
t_{ijk}(\vec{k}_1,\vec{k}_2,\vec{k}_3)\,t_{lmn}(\vec{k}_1,\vec{k}_2,\vec{k}_3)\,\,, 
\end{equation} 
where $\smash{t_{ijk}(\vec{k}_1,\vec{k}_2,\vec{k}_3) = k_1^k\delta_{ij} + k_2^i\delta_{jk} + k_3^j\delta_{ki}}$ 
is symmetric under $\smash{(i\leftrightarrow j,\vec{k}_1 \leftrightarrow\vec{k}_2),\dots}$ 
and has dimension of momentum. One could try to find a generalization of this for the bispectrum from 
$S_{\Lambda_3}$ by writing the amplitude as 
\begin{equation} 
\label{eq:S_Lambda_3_in_in-C} 
A_{S_{\Lambda_3}}(1^{s_1},2^{s_2},3^{s_3})\propto 
\epsilon^{s_1}_{il}({-\vec{k}_1})\epsilon^{s_2}_{jm}({-\vec{k}_2})\epsilon^{s_3}_{kn}({-\vec{k}_3})\, 
\epsilon_{abc}\,T_{lja}(\vec{k}_1,\vec{k}_2)\,T_{mkb}(\vec{k}_2,\vec{k}_3)\,T_{nic}(\vec{k}_3,\vec{k}_1)\,\,, 
\end{equation} 
where $\smash{T_{ijk}(\vec{k},\vec{q})}$ has again dimension of momentum, is antisymmetric in the 
simultaneous interchange of the first two indices and of the two arguments, and does not contain 
the Levi-Civita symbol (that has been factorized outside). Here we have used the conservation of 
momentum to write $\smash{T_{ijk}}$ as a function of two momenta only: unlike \eq{S_Lambda_3_in_in-B}, 
this should not lead to any loss of generality since we have three powers of $\smash{T_{ijk}}$ in \eq{S_Lambda_3_in_in-C}. 

It is however very cumbersome to check this from the the expression in Appendix~\ref{app:cubic_action_3DCS}, 
which contains $\smash{42}$ terms after symmetrization. For this reason we do not pursue this further, 
and only report a version of $\smash{A_{S_{\Lambda_3}}(1^{s_1},2^{s_2},3^{s_3})}$ specialized 
to the polarization tensors $\smash{\epsilon^{\pm}_{ij}(\vec{k})}$. We find 
\begin{equation}
\label{eq:S_Lambda_3_in_in-D}
A_{S_{\Lambda_3}}(1^{s_1},2^{s_2},3^{s_3}) = \sum_{a=1}^6 A^{(a)}_{S_{\Lambda_3}}(1^{s_1},2^{s_2},3^{s_3})\,\,, 
\end{equation} 
where, using the shorthand $\smash{\epsilon^\ast_s}$ for $\smash{\epsilon^s_{ij}({-\vec{k}}) = \epsilon^s_{ij}(\vec{k})^\ast}$, we have 
\begin{subequations}
\label{eq:S_Lambda_3_in_in-E}
\begin{align}
A^{(1)}_{S_{\Lambda_3}}(1^{s_1},2^{s_2},3^{s_3}) &= \frac{1}{4} 
\Tr\big(\epsilon_{s_1}^\ast\cdot\epsilon_{s_2}^\ast\cdot\epsilon_{s_3}^\ast\big) 
\,\big((\vec{k}_1\cdot\vec{k}_2)\,(\lambda_{s_1}k_1 + \lambda_{s_2}k_2) + \text{2 perms.}\big)\,\,, \label{eq:S_Lambda_3_in_in-E-1} \\
A^{(2)}_{S_{\Lambda_3}}(1^{s_1},2^{s_2},3^{s_3}) &= \frac{1}{4} 
\Tr\big(\epsilon_{s_2}^\ast\cdot\epsilon_{s_3}^\ast\big) 
\,(\vec{k}_3\cdot\epsilon^\ast_{s_1}\cdot\vec{k}_3)\,(\lambda_{s_2}k_2 + \lambda_{s_3}k_3) 
+ \text{2 perms.}\,\,, \label{eq:S_Lambda_3_in_in-E-2} \\ 
A^{(3)}_{S_{\Lambda_3}}(1^{s_1},2^{s_2},3^{s_3}) &= {-\frac{1}{2}}\lambda_{s_1}k_1\, 
(\vec{k}_2\cdot\epsilon^\ast_{s_3}\cdot\epsilon^\ast_{s_1}\cdot\epsilon^\ast_{s_2}\cdot\vec{k}_3) 
+ \text{2 perms.}\,\,, \label{eq:S_Lambda_3_in_in-E-3} \\ 
A^{(4)}_{S_{\Lambda_3}}(1^{s_1},2^{s_2},3^{s_3}) &= \frac{\lambda_{s_1}}{6}\Big( 
(\vers{k}_1\cdot\epsilon^\ast_{s_2}\cdot\vec{k}_1)\,(\vec{k}_1\cdot\epsilon^\ast_{s_3}\cdot\epsilon^\ast_{s_1}\cdot\vec{k}_2) 
\nonumber \\ 
&\;\;\;\;\hphantom{\frac{\lambda_{s_1}}{6}\Big(} + (\vers{k}_1\cdot\epsilon^\ast_{s_3}\cdot\vec{k}_1)\, 
(\vec{k}_1\cdot\epsilon^\ast_{s_2}\cdot\epsilon^\ast_{s_1}\cdot\vec{k}_3)\Big) + \text{2 perms.}\,\,, \label{eq:S_Lambda_3_in_in-E-4} \\ 
A^{(5)}_{S_{\Lambda_3}}(1^{s_1},2^{s_2},3^{s_3}) &= {-\frac{1}{2}}(\lambda_{s_1}k_1+\lambda_{s_2}k_2)\, 
(\vec{k}_2\cdot\epsilon^\ast_{s_1}\cdot\epsilon^\ast_{s_3}\cdot\epsilon^\ast_{s_2}\cdot\vec{k}_1) 
+ \text{2 perms.}\,\,, \label{eq:S_Lambda_3_in_in-E-5} \\ 
A^{(6)}_{S_{\Lambda_3}}(1^{s_1},2^{s_2},3^{s_3}) &= \frac{\lambda_{s_3}}{4}(\vec{k}_1\cdot\vec{k}_2) 
\Big(2\vers{k}_3\cdot\epsilon^\ast_{s_1}\cdot\epsilon^\ast_{s_3}\cdot\epsilon^\ast_{s_2}\cdot\vec{k}_3 
- (\vers{k}_3\cdot\epsilon^\ast_{s_2}\cdot\vec{k}_3)\,\Tr(\epsilon^\ast_{s_1}\cdot\epsilon^\ast_{s_3}) \nonumber \\
&\;\;\;\;\hphantom{\frac{\lambda_{s_3}}{4}(\vec{k}_1\cdot\vec{k}_2)\Big( } 
- (\vers{k}_3\cdot\epsilon^\ast_{s_1}\cdot\vec{k}_3)\,\Tr(\epsilon^\ast_{s_2}\cdot\epsilon^\ast_{s_3})\Big) 
+ \text{2 perms.}\,\,, \label{eq:S_Lambda_3_in_in-E-6} 
\end{align} 
\end{subequations} 
where permutations always follow the pattern $\smash{1\to2\to3\to1}$. 
To obtain these expressions we have repeatedly used the relations $\smash{\epsilon_{ijk} k^j\epsilon^s_{ak}({-\vec{k}}) = 
\iu\lambda_s k\epsilon^s_{ai}({-\vec{k}})}$ and its ``inverse'' 
$\smash{\epsilon^s_{ai}({-\vec{k}}) = {-\iu}\lambda_s\epsilon_{ijk}\hat{k}^j\epsilon^s_{ak}({-\vec{k}})}$, 
together with conservation of momentum and the relation $\smash{k^i\epsilon_{ij}^s({-\vec{k}})} = 0$. 

Let us briefly comment on this result in light of \cite{Goodhew:2020hob,Pajer:2020wxk}. 
Studying the limit $\smash{k_T\to 0}$ (intended as limit in the space of complex 
$\smash{k_i}$), these works show how after ``trimming'' the amplitude discussed above from the tree-level bispectrum of 
an operator of order $p$ in derivatives, the result is a function of $\smash{k_1k_2k_3}$ 
and $\smash{k_T}$ that has a pole $\smash{1/k_T^p}$. 
From the point of view of in-in calculations, this can be tracked to the fact that each derivative 
comes with a factor of $\smash{1/a=-H\eta}$ (either from $\smash{\dif/\dif t = \dif/a\dif\eta}$ 
or from the fact that spatial derivatives are contracted with the inverse metric, that is $\smash{\propto 1/a^2}$), 
see e.g.~Section~2.3 of \cite{Pajer:2020wxk} for a discussion. We see that our \eq{S_Lambda_3_in_in-A} does not manifestly 
show such a pole $\smash{1/k_T^3}$, as expected from an operator of this order in derivatives (from \eq{S_Lambda_1} we 
see that this is instead manifest for $\smash{S_{\Lambda_1}}$). 

It would be interesting to investigate this further. We notice that 
operators that break parity involve the volume form $\smash{\vec{e}^{\mu\nu\rho\sigma}}\sim 1/a^3\sim\eta^3$: 
this na{\" i}vely seems to make the above counting in derivatives more complicated. We do not think that this should 
be a problem since the results of \cite{Goodhew:2020hob,Pajer:2020wxk} do not make any assumption about the 
transformation of interactions under parity. Moreover, from the point of view of the in-in calculation itself, 
we see that the operator of $\smash{S_{\Lambda_3}}$ is written in terms of the Christoffel symbols of the spatial metric. 
These do not carry powers of $\smash{a}$: hence, it is the volume form that gives the same overall $\smash{1/a^3}$ as 
for the operator of $\smash{S_{\Lambda_1}}$ (where it comes from three time derivatives).\footnote{This 
happens also in the operator of $\smash{S_{\Lambda_2}}$: in \eq{three_quadratic_parity_odd-1} we have 
$\smash{D_\nu\sim a^0}$, $\smash{\delta\!K_{\rho\lambda}\sim a}$ and $\smash{\delta\!K^\lambda_{\hp{\lambda}\sigma}\sim 1/a}$.} 
A definitive check, i.e.~in order to see if somehow 
\eqsIV{S_Lambda_3_in_in-A}{S_Lambda_3_in_in-C}{S_Lambda_3_in_in-D}{S_Lambda_3_in_in-E} 
``hide'' a pole $\smash{1/k_T^3}$, would be to do a brute-force computation of the limit $\smash{k_T\to 0}$. This becomes 
extremely cumbersome unless one uses more advanced techniques like the spinor helicity formalism. We leave this to future work. 

What about the mixed correlation functions involving scalars? They vanish identically: indeed, 
integrating by parts and using the antisymmetry of $\epsilon_{ijk}$ one can show that 
the vertices $S_{\Lambda_3}|_{\zeta\gamma\gamma}$ and $S_{\Lambda_3}|_{\gamma\zeta\zeta}$ are zero. 
Therefore $\braket{\ggz}$ and $\braket{\gzz}$ are not enhanced in slow roll with respect to Maldacena's ones.

\section{Operators at NNLO in derivatives}
\label{sec:NNLO}

\noindent In this section we play the same game as in 
Section~\ref{sec:four_operators} but focusing on 
operators at fourth order in derivatives (see also 
\cite{Bartolo:2020gsh} for a recent study of these operators).

\subsection{Parity-even operators}
\label{subsec:NNLO_parity_even}

\noindent First, let us consider parity-even operators that start quadratic in perturbations. 
These have been identified in \cite{Bordin:2017hal}: we recap them here for convenience of the reader. 
Working with the graviton $\gamma_{ij}$, the only structures we can write 
are $\ddot{\gamma}_{ij}\ddot{\gamma}_{ij}$, $\dot{\gamma}_{ij}\partial^2\dot{\gamma}_{ij}$ 
and $\partial^2\gamma_{ij}\partial^2\gamma_{ij}$. They correspond to 
\begin{subequations}
\label{eq:NNLO-A}
\begin{align}
\ddot{\gamma}_{ij}\ddot{\gamma}_{ij} &\sim 
n^\rho\nabla_\rho\delta\!K^{\mu\nu} n^\sigma\nabla_\sigma\delta\!K_{\mu\nu}\,\,, \label{eq:NNLO-A-1} \\
\dot{\gamma}_{ij}\partial^2\dot{\gamma}_{ij} &\sim 
{{^{(3)}}\!R_{\mu\nu}}\,n^\rho\nabla_\rho\delta\!K^{\mu\nu}, \ 
\delta\!K^{\mu\nu}D^\rho D_\rho\delta\!K_{\mu\nu}\,\,, \label{eq:NNLO-A-2} \\
\partial^2\gamma_{ij}\partial^2\gamma_{ij} &\sim {{^{(3)}}\!R^{\mu\nu}}{{^{(3)}}\!R_{\mu\nu}}, \ 
{{^{(3)}}\!R_{\rho\sigma\mu\nu}}{{^{(3)}}\!R^{\rho\sigma\mu\nu}}\,\,, \label{eq:NNLO-A-3} 
\end{align}
\end{subequations}
where numerical factors and scale factors are neglected for simplicity. We then move to operators starting cubic in perturbations. 
There are only two structures that we can write, i.e.~$\dot{\gamma}_{ij}\dot{\gamma}_{jk}\partial^2\gamma_{ki}$ 
and $\dot{\gamma}_{ij}\dot{\gamma}_{kl}\partial_k\partial_l\gamma_{ij}$. The corresponding operators are 
\begin{subequations}
\label{eq:NNLO-B}
\begin{align}
\dot{\gamma}_{ij}\dot{\gamma}_{jk}\partial^2\gamma_{ki} &\sim 
{{^{(3)}}\!R_{\mu\nu}}\delta\!K^\mu_{\hp{\mu}\rho}\delta\!K^{\rho\nu}\,\,, \label{eq:NNLO-B-2} \\
\dot{\gamma}_{ij}\dot{\gamma}_{kl}\partial_k\partial_l\gamma_{ij} &\sim 
{{^{(3)}}\!R_{\rho\sigma\mu\nu}}\delta\!K^{\rho\mu}\delta\!K^{\sigma\nu}\,\,. \label{eq:NNLO-B-3} 
\end{align}
\end{subequations}

\subsection{Integration by parts and field redefinitions}
\label{subsec:NNLO_IBPs_and_redefinitions}

\noindent Are some of these operators redundant? 
First we consider the operator $\smash{{{^{(3)}}\!R_{\rho\sigma\mu\nu}}{{^{(3)}}\!R^{\rho\sigma\mu\nu}}}$ 
in \eq{NNLO-A-3}. We can rewrite it in terms of lower-derivative operators and the remaining operators in \eqsII{NNLO-A}{NNLO-B} 
using the $3+1$ decomposition of the Riemann tensor (summarized in Appendix~\ref{app:appendix-C}) and the fact that the 
Gauss-Bonnet combination of \eq{GB-1} is a total derivative. Then, in Appendix~\ref{app:appendix-0} we show that 
\begin{equation}
\label{eq:IBP_at_NNLO}
\begin{split}
{{^{(3)}}\!R_{\mu\nu}}n^\rho\nabla_\rho K^{\mu\nu} 
&= 2D_\sigma K^{\rho\sigma} D_\mu K^\mu_{\hp{\mu}\rho} 
+ 2{{^{(3)}}\!R^\rho_{\hp{\rho}\lambda\mu\sigma}}K^{\lambda\sigma}K^\mu_{\hp{\mu}\rho} + K^{\rho\sigma}D^\mu D_\mu K_{\rho\sigma} \\
&\;\;\;\; + \text{boundary terms} + \text{lower-derivative operators} \\
&\;\;\;\; + \text{operators involving scalar modes}\,\,,
\end{split}
\end{equation}
where the first term on the right-hand side starts at fourth order in $\gamma_{ij}$. This tells us 
that it is sufficient to consider $\smash{{{^{(3)}}\!R_{\mu\nu}}\,n^\rho\nabla_\rho\delta\!K^{\mu\nu}}$ in the action up 
to a redefinition of the coefficients of the cubic operator of \eq{NNLO-B-3} and of lower-derivative operators. 

Let us discuss the field redefinitions. Starting at linear level in fluctuations we have only 
\begin{equation}
\label{eq:redefinitions_NNLO-A}
g^{\mu\nu}\to g^{\mu\nu} + c_{{{^{(3)}}\!R}}{{^{(3)}}\!R^{\mu\nu}} 
+ c_{\dot{K}}n^\rho\nabla_\rho\delta\!K^{\mu\nu}\,\,, 
\end{equation}
where $\smash{c_{{{^{(3)}}\!R}}}$ and $\smash{c_{\dot{K}}}$ have the dimensions 
of a length squared. Using the projection of the Einstein tensor on the hypersurfaces of constant time, i.e.~ 
\begin{equation}
\label{eq:redefinitions_NNLO-B}
\begin{split}
h_\sigma^{\hp{\sigma}\gamma} h_\nu^{\hp{\nu}\delta}G_{\gamma\delta} &= {{^{(3)}}\!R_{\mu\nu}} + K K_{\sigma\nu} 
+ n^\delta\nabla_\delta K_{\sigma\nu} - \frac{1}{2}{{^{(3)}}\!R}\,h_{\sigma\nu} + \frac{1}{2}K^2\,h_{\sigma\nu} 
- \frac{1}{2} K_{\alpha\beta}K^{\alpha\beta}\,h_{\sigma\nu} \\ 
&\;\;\;\; - h_{\sigma\nu}\nabla_\rho(Kn^\rho) + \text{terms involving $A^\mu$}\,\,,
\end{split}
\end{equation}
we see that the variation of the Einstein-Hilbert action under \eq{redefinitions_NNLO-A} contains 
the three operators $\smash{n^\rho\nabla_\rho\delta\!K^{\mu\nu} n^\sigma\nabla_\sigma\delta\!K_{\mu\nu}}$, 
$\smash{\delta\!K^{\mu\nu}n^\rho\nabla_\rho{{^{(3)}}\!R_{\mu\nu}}}$ and 
$\smash{{{^{(3)}}\!R^{\mu\nu}}{{^{(3)}}\!R_{\mu\nu}}}$. We choose to remove the first two: 
this is simply because the last one is easier to deal with once scalar modes are considered since it does not contain the constraints. 

Let us conclude this section by studying the field redefinitions for the cubic operators. 
It is straightforward to see that we can generate the operator 
$\smash{{{^{(3)}}\!R_{\mu\nu}}\delta\!K^\mu_{\hp{\mu}\rho}\delta\!K^{\rho\nu}}$ 
from the first term in \eq{redefinitions_NNLO-B} with the field redefinition 
\begin{equation}
\label{eq:NNLO-C}
g^{\mu\nu}\to g^{\mu\nu} + c_{K^2}\delta\!K^\mu_{\hp{\mu}\rho}\delta\!K^{\rho\nu}\,\,.
\end{equation}
On the other hand, it is not possible to generate the operator $\smash{
{{^{(3)}}\!R_{\rho\sigma\mu\nu}}\delta\!K^{\rho\mu}\delta\!K^{\sigma\nu}}$ with any field redefinition 
since the Riemann tensor carries too many indices. We then conclude that at fourth order in derivatives 
there are only two parity-even operators that are not redundant, and only one of them modifies the tensor power spectrum.

\subsection{Parity-odd operators}
\label{subsec:NNLO_parity_odd}

\noindent Let us now move to operators that break parity. At quadratic order in perturbations we have 
only two structures, i.e.~$\epsilon_{ijk}\partial_i\dot{\gamma}_{jl}\ddot{\gamma}_{lk}$ and 
$\epsilon_{ijk}\partial_i\dot{\gamma}_{jl}\partial^2\gamma_{lk}$. 
At the covariant level, these correspond to the operators 
\begin{subequations}
\label{eq:NNLO-D}
\begin{align}
\epsilon_{ijk}\partial_i\dot{\gamma}_{jl}\ddot{\gamma}_{lk} &\sim 
\frac{\vec{e}^{\mu\nu\rho\sigma}n_\mu}{N}\,D_\nu\delta\!K_{\rho\lambda}n^\delta\nabla_\delta
\delta\!K^\lambda_{\hp{\lambda}\sigma}\,\,, \label{eq:NNLO-D-1} \\
\epsilon_{ijk}\partial_i\dot{\gamma}_{jl}\partial^2\gamma_{lk} &\sim 
\frac{\vec{e}^{\mu\nu\rho\sigma}n_\mu}{N}\,D_\nu\delta\!K_{\rho\lambda}{^{(3)}}\!R^\lambda_{\hp{\lambda}\sigma}\,\,. \label{eq:NNLO-D-2}
\end{align}
\end{subequations} 

There is only one option when we move to operators starting cubic in perturbations. If we want to remain with four derivatives the 
antisymmetry of $\epsilon_{ijk}$ forbids us to use the three-Ricci tensor. Moreover, 
we still cannot use $\smash{{^{(3)}}\!R_{\rho\sigma\mu\nu}}$ since we would remain with a free index on $\epsilon_{ijk}$ 
unless we introduce scalar modes. The only structure we can build is 
$\epsilon_{ijk}\dot{\gamma}_{il}\dot{\gamma}_{lm}\partial_j\dot{\gamma}_{mk}$. At the covariant level this is 
\begin{equation}
\label{eq:NNLO-E}
\epsilon_{ijk}\dot{\gamma}_{il}\dot{\gamma}_{lm}\partial_j\dot{\gamma}_{mk}\sim 
\frac{\vec{e}^{\mu\nu\rho\sigma}n_\mu}{N}\,\delta\!K_\nu^{\hp{\nu}\lambda} 
\delta\!K_\lambda^{\hp{\lambda}\gamma} D_\rho\delta\!K_{\gamma\sigma}\,\,. 
\end{equation}

Let us first discuss integration by parts. It is straightforward to see that 
the cubic operator of \eq{NNLO-E} is not a total derivative. 
On the other hand, the two operators of \eqsI{NNLO-D} are total divergences at quadratic order in $\gamma_{ij}$. 
In Appendix~\ref{app:appendix-0} we show that we can remove the operator of \eq{NNLO-D-1} at all 
orders in perturbations, up to a redefinition of the coefficients of lower-derivative operators and of the operator of \eq{NNLO-E}. 

One might wonder if the same holds for the quadratic operator involving the three-Ricci tensor, cf.~\eq{NNLO-D-2}. 
We have a reason to believe that this is not the case. Indeed, showing that the operator is a 
total divergence at quadratic order in $\gamma_{ij}$ explicitly 
requires to use the linear-order relation ${{^{(3)}}\!R_{ij}} = -\partial^2\gamma_{ij}/2$, that 
does not have an analogue at the nonperturbative level. 
This is only a hint: a definitive proof would be, e.g., computing 
the equations of motion for this operator (see \cite{tHooft:1974toh}, for example). 
In the rest of this paper we take this to be an operator that starts cubic in $\gamma_{ij}$.\footnote{Notice 
that there are other cases of operators that are total divergences at a finite order in perturbations 
but not at all orders. One example is $\smash{{{^{(3)}}\!R}}$.} 

We conclude by discussing field redefinitions. In order to remove these cubic 
operators we would need a symmetric tensor involving $\vec{e}^{\mu\nu\rho\sigma}$ 
that starts quadratic in perturbations and carries two derivatives acting on the metric. Similarly to what happened 
at third order in derivatives, it is not possible to build such a tensor.

\subsection{Three-point functions involving scalars}
\label{subsec:three_ricci_tensor_squared}

\noindent In this section we compute the scalar-tensor-tensor and tensor-scalar-scalar three-point 
functions from $\smash{{{^{(3)}}\!R_{\mu\nu}}{{^{(3)}}\!R^{\mu\nu}}}$. The action is 
\begin{equation} 
\label{eq:last_operator} 
S = S_0 + M^2_{\rm P}\int\dif^4x\,\sqrt{-g}\,\frac{{{^{(3)}}\!R_{\mu\nu}}{{^{(3)}}\!R^{\mu\nu}}}{\widebar{\Lambda}^2_1}\,\,. 
\end{equation} 
This operator gives a correction to the power spectrum of the graviton 
proportional to $\smash{H^2/\widebar{\Lambda}^2_1}$, as shown in \cite{Bordin:2017hal}. 
Similarly to the discussion in Section~\ref{subsec:estimates}, we expect that the ratio of 
the $\braket{\ggz}$ and $\braket{\gzz}$ three-point functions with those coming from $S_0$ 
is enhanced by $1/\varepsilon$. In fact, the explicit calculation reveals that 
this holds only for the former: the latter vanishes at late times. 

Since $\smash{{{^{(3)}}\!R_{\mu\nu}}{{^{(3)}}\!R^{\mu\nu}}}$ does not contain the lapse and shift constraints, 
being built exclusively from the connection coefficients of the three-dimensional covariant derivative, 
it is easier to carry out the calculation in $\zeta$ gauge. The resulting three-point function is\footnote{The 
computation in \cite{Bordin:2017hal} was missing the piece $\smash{\delta\!N\,{{^{(3)}}\!R_{\mu\nu}}{{^{(3)}}\!R^{\mu\nu}}}$ 
in the cubic action ($\delta\!N = \dot{\zeta}/H$), which gives the first term in \eq{ggz_last_operator-A}.} 
\begin{equation}
\label{eq:ggz_last_operator-A}
\begin{split}
\Delta\!\braket{\zeta_{\vec{k}_1}\gamma^{s_2}_{\vec{k}_2}\gamma^{s_3}_{\vec{k}_3}}' 
&= \frac{H^2}{\widebar{\Lambda}^2_1}\frac{H^4}{8\eps\mpl^4}\frac{k^2_1+5k_1(k_2+k_3)+4(k^2_2+5k_2k_3+k^2_3)}{k_1k_2k_3 k_T^5}\, 
\Tr(\epsilon^\ast_{s_2}\cdot\epsilon^\ast_{s_3}) \\
&\;\;\;\; + \frac{H^2}{\widebar{\Lambda}^2_1}\frac{H^4}{4\eps\mpl^4} 
\frac{\widebar{{\cal I}}_a(1,2,3)}{k_1^3k_2^3k_3^3 k_T^4} 
\Big(\widebar{{\cal I}}_b(1,2^{s_2},3^{s_3}) + \widebar{{\cal I}}_c(1,2^{s_2},3^{s_3})\Big)\,\,, 
\end{split}
\end{equation} 
where 
\begin{equation}
\label{eq:ggz_last_operator-B}
\begin{split}
\widebar{{\cal I}}_a(1,2,3) = k^3_1 + k^3_2 + k^3_3 + 4k^2_2k_3 
+ 4 k_2 k^2_3 + 4k^2_1(k_2+k_3) + 4k_1(k^2_2+3k_2k_3+k^2_3)\,\,, 
\end{split}
\end{equation}
\begin{equation}
\label{eq:ggz_last_operator-C}
\begin{split}
\widebar{{\cal I}}_b(1,2^{s_2},3^{s_3}) = 
\bigg(\frac{k^2_3\,\vec{k}_1\cdot\vec{k}_2 + k^2_2\,\vec{k}_1\cdot\vec{k}_3}{2} + k^2_1\,\vec{k}_2\cdot\vec{k}_3 + 
(\vec{k}_1\cdot\vec{k}_2)(\vec{k}_1\cdot\vec{k}_3) - \frac{k^2_2k^2_3}{2}\bigg) 
\Tr(\epsilon^\ast_{s_2}\cdot\epsilon^\ast_{s_3}) 
\end{split}
\end{equation} 
and 
\begin{equation}
\label{eq:ggz_last_operator-D}
\begin{split}
\widebar{{\cal I}}_c(1,2^{s_2},3^{s_3}) &= (\vec{k}_1\cdot\epsilon^\ast_{s_2}\cdot\vec{k}_1) 
(\vec{k}_1\cdot\epsilon^\ast_{s_3}\cdot\vec{k}_1)\,\,. 
\end{split} 
\end{equation} 
Here we have repeatedly used the transverseness condition together with the 
conservation of momentum to simplify the tensor structure as much as possible.

\subsection[About the enhancement by \texorpdfstring{$1/\varepsilon$}{1/\textbackslash epsilon}]
{About the enhancement by \texorpdfstring{$\vec{1/\varepsilon}$}{1/\textbackslash epsilon}}
\label{subsec:about_enhancement}

\noindent Before proceeding, it is important to discuss in more detail the $\smash{1/\varepsilon}$ 
enhancement of the above correlator. The operator of \eq{last_operator} modifies also the scalar 
power spectrum (unlike the one of \eq{three_quadratic_parity_odd-2} at NLO in derivatives), since 
$\smash{{{^{(3)}}\!R_{ij}}}$ contains $\smash{\delta_{ij}\partial^2\zeta}$ and $\smash{\partial_i\partial_j\zeta}$. 
It is straightforward to see that the fractional correction to the power spectrum of $\zeta$ is of order 
$\smash{(H^2/\widebar{\Lambda}^2_1)\,\varepsilon^{-1}}$. This tells us that if we take 
$\smash{(H^2/\widebar{\Lambda}^2_1)\,\varepsilon^{-1}\gg 1}$ the corrections to the 
the scalar mode functions become dominant, and our calculation of the three-point function 
using de Sitter modes does not make sense. 

Let us then consider the combination 
\begin{equation}
\label{eq:about_enhancement}
{{^{(3)}}\!R_{\mu\nu}}{{^{(3)}}\!R^{\mu\nu}}\to {{^{(3)}}\!R_{\mu\nu}}{{^{(3)}}\!R^{\mu\nu}} - \frac{3}{8}{{^{(3)}}\!R^2} 
\end{equation} 
in \eq{last_operator}. It vanishes at quadratic order in $\smash{\zeta}$ after integration by parts. 
For this operator, then, we have that $\smash{(H^2/\widebar{\Lambda}^2_1)\,\varepsilon^{-1}}$ 
is proportional to $\smash{M^2_{\rm P}A_{\rm s}/\widebar{\Lambda}^2_1}$, and it makes sense to consider 
the limit $\smash{M^2_{\rm P}A_{\rm s}/\widebar{\Lambda}^2_1\gg 1}$ in which $\smash{\braket{\ggz}_{\widebar{\Lambda}^2_1}}$ 
dominates over $\smash{\braket{\ggz}_{0}}$. The contribution of the $\smash{{{^{(3)}}\!R^2}}$ term to 
the scalar-tensor-tensor bispectrum of \eq{ggz_last_operator-A} is to send 
\begin{equation}
\label{eq:three_ricci_scalar_squared}
\widebar{{\cal I}}_b(1,2^{s_2},3^{s_3}) \to \widebar{{\cal I}}_b(1,2^{s_2},3^{s_3}) 
- \frac{3}{2} k^2_1\,(\vec{k}_2\cdot\vec{k}_3)\,\Tr(\epsilon^\ast_{s_2}\cdot\epsilon^\ast_{s_3}) 
\end{equation} 
in \eq{ggz_last_operator-C}. 

Importantly, the operator of \eq{about_enhancement} still contributes 
to $\smash{\braket{\zzz}}$: the corresponding $\smash{\fnl{\zzz}}$ is 
of order $\smash{(H^2/\widebar{\Lambda}^2_1)\,\varepsilon^{-1}\sim 
M^2_{\rm P}A_{\rm s}/\widebar{\Lambda}^2_1}$. Given the current constraints 
on non-Gaussianities not of local type, there is however still parameter space 
where this quantity can be $\smash{\gg 1}$ while still having $\smash{\widebar{\Lambda}_1}$ 
lower than the Planck scale and $\smash{H^2/\widebar{\Lambda}^2_1}$ not too close to $\smash{1}$ 
(so that the EFT expansion is under control).\footnote{We also stress that, in case of detection 
of the particular combination of \eq{about_enhancement}, one should address the tuning between the two operators.}

\section{About the consistency relations} 
\label{sec:CRs}

\noindent In this section we discuss in more detail the soft limits of the three-point functions 
of Sections~\ref{sec:NGs} and~\ref{sec:NNLO}. Let us start from Section~\ref{subsec:three_ricci_tensor_squared}. 
Since we our power spectra are scale-invariant, we expect that the three-point function of \eq{ggz_last_operator-A} 
vanishes up to order $\smash{1/q}$, where $\smash{\vec{q}}$ is the momentum of the long-wavelength scalar mode. 
It is straightforward to see that this does not happen: the last term in \eq{ggz_last_operator-C} 
contributes at order $\smash{1/q^3}$. 

Let us see why this happens. The three-point function discussed in Section~\ref{subsec:three_ricci_tensor_squared} 
is only the contribution from the cubic scalar-tensor-tensor vertex contained in $\smash{{{^{(3)}}\!R_{\mu\nu}} 
{{^{(3)}}\!R^{\mu\nu}}}$. This operator, however, modifies also the quadratic action for the scalar mode. At tree level, 
there is then also a diagram in which the cubic vertex is the one from the Einstein-Hilbert action, but there is a 
$\smash{\zeta\zeta}$ ``mixing'': this contributes at the same order in $\smash{1/\widebar{\Lambda}^2_1}$. 

These additional diagrams are not the solution, however: once we consider the combination of \eq{about_enhancement}, 
the quadratic scalar action is not modified, but one can see that the $\smash{1/q^3}$ term still survives since 
\eq{three_ricci_scalar_squared} only contributes at order $\smash{1/q}$ in the squeezed limit. 

The solution of the puzzle is the fact that the cubic scalar-tensor-tensor vertex in the Einstein-Hilbert action 
is suppressed by $\smash{\varepsilon}$ only after a field redefinition. More precisely, one finds \cite{Maldacena:2002vr} 
\begin{equation}
\label{eq:CRs-A}
S_0\supset{-\int\dif^4x\,\frac{\zeta}{H}\dot{\gamma}_{ij}\frac{\partial{\cal L}_{0,\gamma\gamma}}{\partial\gamma_{ij}}}\,\,,
\end{equation}
where $\smash{{\cal L}_{0,\gamma\gamma}}$ is the quadratic Lagrangian for the graviton from $S_0$, 
which is not suppressed by $\smash{\varepsilon}$. This term is removed in \cite{Maldacena:2002vr} by a field 
redefinition $\smash{\gamma_{ij}\to\gamma_{ij} + \zeta\dot{\gamma}_{ij}/H}$. Things are slightly more tricky 
now, since the quadratic action for $\smash{\gamma_{ij}}$ is modified by $\smash{{{^{(3)}}\!R_{\mu\nu}} 
{{^{(3)}}\!R^{\mu\nu}}}$. Essentially, the field redefinition leaves us with the additional cubic vertex 
\begin{equation}
\label{eq:CRs-B}
\frac{M^2_{\rm P}}{2\widebar{\Lambda}^2_1}\int\dif^4x\,\frac{\zeta}{aH}\dot{\gamma}_{ij}\partial^4\gamma_{ij}\,\,. 
\end{equation} 
This gives a contribution 
\begin{equation}
\label{eq:CRs-C}
\frac{H^2}{\widebar{\Lambda}^2_1}\frac{H^4}{8\eps\mpl^4}\frac{\widebar{{\cal I}}_{\rm redef}(1,2,3)}{k_1^3k_2^3k_3^3 k_T^5}\, 
\Tr(\epsilon^\ast_{s_2}\cdot\epsilon^\ast_{s_3})\,\,, 
\end{equation} 
where 
\begin{equation}
\label{eq:CRs-D}
\widebar{{\cal I}}_{\rm redef}(1,2,3) = k^2_2k^4_3\big(4k^2_1 + k^2_2 + 5k_2k_3 + 4k^2_3 + 5k_1(k_2+4k_3)\big) + (2\to 3)\,\,. 
\end{equation} 
It is straightforward to check that its soft limit exactly cancels the $\smash{1/q^3}$ term in \eq{ggz_last_operator-A}. 

What about the tensor-tensor-tensor three-point function from the three-dimensional Chern-Simons term? 
A long-wavelength adiabatic tensor mode is equivalent to an anisotropic rescaling of coordinates 
at leading order in gradients. Hence it has an effect also on scale-invariant power spectra. Since 
$\smash{S_{\Lambda_3}}$ modifies the graviton power spectrum, we expect that at order $\smash{H/\Lambda_3}$ 
we have a contribution of order $\smash{1/q^3}$ to the graviton bispectrum. 

While \eq{S_Lambda_3_in_in-A} does indeed have a term of this order, it is not enough to satisfy the consistency relation 
\begin{equation}
\label{eq:CRs-E}
\braket{\gamma^{s}_{\vec{q}}\gamma^{r}_{\vec{k}-\vec{q}/2}\gamma^{r}_{{-\vec{k}}-\vec{q}/2}}' = 
\frac{3}{2}P_{\gamma^s}(q)\epsilon^s_{ij}(\vec{q})\hat{k}^i\hat{k}^jP_{\gamma^r}(k)\,\,,
\end{equation} 
valid up to fractional corrections of order $\smash{q^2/k^2}$.\footnote{Notice that here we 
have allowed the possibility that the power spectrum of the graviton depends on the helicity, 
as it happens in the case of parity-violating operators like $\smash{S_{\Lambda_3}}$.} 
The reason is that \eq{S_Lambda_3_in_in-A} and related come only from the cubic vertex in 
$\smash{S_{\Lambda_3}}$, and do not account for the contribution of the Einstein-Hilbert 
action plus the $\smash{\gamma\gamma}$ ``mixing'' from the quadratic part of $\smash{S_{\Lambda_3}}$. 
We leave a computation of these additional contributions to future work.

\section{Conclusions} 
\label{sec:conclusions}

\noindent In this paper we studied the perturbative non-Gaussianities of the graviton 
in the framework of single-field inflation. When couplings of the metric to the foliation are allowed, 
deviations from Einstein gravity are present already at next-to-leading order in derivatives. 

At order $H/\Lambda$ the split in the graviton helicities can be traced to a single operator, that is 
the Chern-Simons term on the hypersurfaces of constant clock. This operator modifies the graviton bispectrum, 
but at leading order in slow roll it does not correct the mixed correlators $\braket{\ggz}$ and $\braket{\gzz}$. 
There is an additional operator that breaks parity but does not affect super-horizon correlators (it gives only a correction 
to the phase of the wavefunction of the universe). Finally, an operator starting cubic in metric perturbations 
gives the leading parity-conserving correction to the graviton bispectrum. 

At order $H^2/\Lambda^2$ there are no parity-odd modifications of the power spectrum and a single parity-even one. 
Of the three operators starting cubic in perturbations, one conserves parity and two do not. 
The operator that modifies the power spectrum affects $\braket{\ggz}$ as well: 
this new contribution is more sizable than Maldacena's result if 
the ratio $H^2/\Lambda^2$ is larger than the slow-roll parameter $\varepsilon$ 
(equivalently, using the normalization $A_{\rm s}$ of the scalar power spectrum, if the scale 
$\Lambda$ is much lower than $M_{\rm P}\sqrt{A_{\rm s}}\,$). We argue that 
this makes this operator a prime target if tensor modes are detected. 

In light of this, it would be interesting to see if it is possible to disentangle the signature of 
this operator in $\braket{\ggz}$ from the contributions of the operators with two derivatives. 
Given that the latter effectively come only from the modification of the constraint equations, 
this amounts to study what are the most general solutions for $\delta\!N$ and $N_{\rm L}$ at leading order in derivatives. 

We conclude by emphasizing that the results of this paper provide important theoretical data for the 
``boostless cosmological bootstrap'' program being developed e.g.~in \cite{Pajer:2020wnj,Goodhew:2020hob,Pajer:2020wxk}.

\section*{Acknowledgements}

\noindent We thank Paolo Creminelli, Victor Gorbenko, Austin Joyce, Luca Santoni, 
Fabian Schmidt and especially Maria Alegria Gutierrez, Mehrdad Mirbabayi and Enrico Pajer for useful discussions. 
L.~B. is supported by STFC Consolidated Grant No.~ST/P000703/1. 
G.~C.~acknowledges support from the Starting Grant (ERC-2015-STG 678652) ``GrInflaGal'' from the European Research Council. 
L.~B.~thanks ICTP for hospitality while part of this work was carried out.

\appendix

\section{Notation and conventions}
\label{app:notation}

\noindent In this appendix we summarize our notation and conventions for convenience of the reader. The unitary-gauge line element is 
\begin{equation}
\label{eq:unitary_gauge_line_element}
\dif s^2 = -N^2\dif t^2 + h_{ij}(\dif x^i + N^i\dif t)(\dif x^j + N^j\dif t)\,\,,
\end{equation}
where 
\begin{equation}
\label{eq:gamma_and_zeta}
h_{ij} = a^2\eu^{2\zeta}(\eu^\gamma)_{ij}\,\,,\quad\gamma_{ii} = 0\,\,,\quad\partial_i\gamma_{ij} = 0\,\,.
\end{equation}
We write the shift function as 
\begin{equation}
\label{eq:lapse_and_shift}
N_i = h_{ij}N^j = D_i N_{\rm L} + h_{ij}N^j_{\rm T}\,\,,\quad D_i N^i_{\rm T} = 0\,\,. 
\end{equation} 
Notice that, at linear order in perturbations, this definition has an overall $a^{-2}$ with respect to that of 
\cite{Maldacena:2002vr}. The transverse part of the shift does not play a role in this work 
(vector modes are not excited since we work perturbatively around the minimal single-clock action). 

The coordinate transformation from unitary gauge to flat gauge, i.e.~the reintroduction of the Stueckelberg field $\pi$, 
is defined as $t_\zeta = t_\pi + \pi(t_\pi,\vec{x})$ \cite{Cheung:2007st}: 
we will use the shorthand $t\to t+\pi$ for this transformation. 

Our decomposition of scalar and tensor modes is 
\begin{subequations}
\label{eq:fourier_decomposition}
\begin{align}
\zeta(t,\vec{x}) &= \int\frac{\dif^3k}{(2\pi)^3}\zeta_{\vec{k}}(t)\eu^{\iu\vec{k}\cdot\vec{x}}\,\,, \label{eq:fourier_decomposition-1} \\
\gamma_{ij}(t,\vec{x}) &= \int\frac{\dif^3k}{(2\pi)^3}\sum_{s}\epsilon_{ij}^s(\vec{k})
\gamma^s_{\vec{k}}(t)\eu^{\iu\vec{k}\cdot\vec{x}}\,\,, \label{eq:fourier_decomposition-2}
\end{align}
\end{subequations} 
and similarly for the Stueckelberg field $\smash{\pi(t,\vec{x})}$. Here, the traceless polarization 
tensors $\smash{\epsilon^s_{ij}}$ satisfy $\smash{k^i\epsilon_{ij}^s(\vec{k})} = 0$. 
For the circularly-polarized tensors, the property 
$\smash{\iu k^l\epsilon_{jlm}\epsilon^s_{im}(\vec{k}) 
= \lambda_s k\epsilon^s_{ij}(\vec{k})}$, $\smash{\lambda_\pm = \pm1}$, is also repeatedly used 
(also, we recall that $\smash{\epsilon^s_{ij}({-\vec{k}}) = \epsilon^s_{ij}(\vec{k})^\ast}$). 
One can use the following expression for $\smash{\epsilon^{\pm}_{ij}(\vec{k})}$: 
\begin{equation}
\label{eq:polarization_tensors_higher_spins}
\epsilon^{\pm}_{ij}(\vec{k}) = (\hat{u}_i\hat{u}_j-\hat{v}_i\hat{v}_j)\mp\iu(\hat{u}_i\hat{v}_j+\hat{v}_i\hat{u}_j)\,\,, 
\end{equation}
where $\vers{v}$ and $\vers{u}=\vers{k}\times\vers{v}/\abs{\vers{k}\times\vers{v}}$ are unit vectors orthogonal to $\vec{k}$. 
Our normalization of the tensor power spectrum follows that of 
\cite{Maldacena:2011nz,Creminelli:2014wna}: we have $\smash{\epsilon^s_{ij}(\vec{k})\epsilon^{s'}_{ij}({-\vec{k}}) = 4\delta_{ss'}}$, 
and the tensor power spectrum $\smash{P_{\gamma^s}(k)}$ from the Einstein-Hilbert action is equal to $\smash{H^2/2M^2_{\rm P}k^3}$. 

Primes on correlation functions denote that we have removed a factor of $(2\pi)^3$ times the Dirac delta function of momentum 
conservation, and we will drop the time argument on the Fourier modes when we look at late-time correlation functions. 

The in-in master formula for the vacuum expectation value $\braket{{\cal O}(t)} = 
\braket{\Omega|{\cal O}(t)|\Omega}$ of an operator ${\cal O}(t)$ 
(for example $\smash{{\cal O}(t) = \gamma^s_{\vec{k}}(t)\gamma^{s'}_{\vec{k}'}(t)}$ 
for the graviton power spectrum) is (see e.g.~\cite{BaumannLecturesInflation,Chen:2010xka} for a review) 
\begin{equation} 
\label{eq:in_in_master-A}
\braket{{\cal O}(t)} = \Big\langle\Omega\Big|\Big(T\eu^{-\iu\int_{-\infty(1-\iu\epsilon)}^t\dif t'H_I(t')}\Big)^\dag\, 
{\cal O}(t)\,\Big(T\eu^{-\iu\int_{-\infty(1-\iu\epsilon)}^t\dif t''H_I(t'')}\Big)\Big|\Omega\Big\rangle\,\,, 
\end{equation} 
where $H_I$ is the interaction Hamiltonian and the rotation $-\infty(1-\iu\epsilon)$ projects onto the free vacuum. 
All our calculations of correlation functions will stop at tree level. 
Switching to conformal time and focusing on the late-time limit $\eta\to 0$, the formula above then reduces to 
(dropping the $\smash{(1-\iu\epsilon)}$ for simplicity of notation) 
\begin{equation}
\label{eq:in_in_master-B}
\braket{{\cal O}(0)} = \iu\int_{-\infty}^0\dif\eta\,a(\eta)\,\big\langle0\big|\big[H_I(\eta),{\cal O}(0)\big]\big|0\big\rangle\,\,.
\end{equation}

\section{Relation between \texorpdfstring{$\vec{3{\rm D}}$}{3D} and \texorpdfstring{$\vec{4{\rm D}}$}{4D} Chern-Simons terms} 
\label{app:appendix-minus_1}

\noindent In this appendix we show how to relate the four-dimensional Chern-Simons term to the three-dimensional one and 
the operator of $S_{\Lambda_2}$. For simplicity we focus on tensor modes only. 
First, consider the coordinate basis $\partial^\mu_\alpha$ dual to $\dif x^\alpha_\mu$. If we neglect scalar modes, we have 
\begin{subequations}
\label{eq:appendix-minus_1-A}
\begin{align}
K_{\mu\nu} &= \nabla_\mu n_\nu = -\nabla_\mu\dif x^0_\nu\,\,, \label{eq:appendix-minus_1-A-1} \\
K_\mu^{\hp{\mu}\nu} &= \nabla_\mu n^\nu = \nabla_\mu\partial^\nu_0\,\,. \label{eq:appendix-minus_1-A-2}
\end{align}
\end{subequations}
In components, we then find that $\Gamma^0_{\mu\nu} = K_{\mu\nu}$ and $\Gamma^\rho_{\mu 0} = K_\mu^{\hp{\mu}\rho}$. 
Using the fact that for a projected tensor the upper temporal indices vanish, and the lower temporal indices vanish as well 
if we put the shift vector to zero, we find that $\Gamma^0_{0\nu} = 0$ and $\Gamma^\rho_{00} = 0$. 
Finally, with similar manipulations one can show that $\smash{\Gamma^k_{ij} = \threeG{k}{i}{j}}$. 

Using $\vec{e}^{0ijk} = {-\epsilon_{ijk}}/\sqrt{-g}$, in \eq{three_quadratic_parity_odd-3} we have 
\begin{equation}
\label{eq:appendix-minus_1-B}
S_{\Lambda_4} = {-M^2_{\rm P}}\int\dif^4x\, 
\frac{1}{\Lambda_4}\,\epsilon_{ijk}\Bigg(\frac{\Gamma^\sigma_{i\nu}\partial_j\Gamma^\nu_{k\sigma}}{2} 
+ \frac{\Gamma^\sigma_{i\nu}\Gamma^\nu_{j\lambda}\Gamma^\lambda_{k\sigma}}{3}\Bigg)\,\,.
\end{equation} 
Dropping the overall constant ${-M^2_{\rm P}}/\Lambda_4$ for simplicity, and 
expanding the Einstein summation, we isolate the three-dimensional Chern-Simons term plus five additional terms. 
Three of these terms involve two powers of the extrinsic curvature and one of the connection coefficients of the three-dimensional 
covariant derivative. Using the antisymmetry of $\epsilon_{ijk}$, they add to give 
\begin{equation}
\label{eq:appendix-minus_1-C}
{-\int}\dif^4x\,\epsilon_{ijk}K_i^{\hp{i}l}\threeG{m}{j}{l}K_{mk}\,\,.
\end{equation}
The two remaining terms (that involve the extrinsic curvature and its spatial derivative) 
are equal to each other after integration by parts. Their sum is 
\begin{equation}
\label{eq:appendix-minus_1-D}
\int\dif^4x\,\epsilon_{ijk}K_i^{\hp{i}l}\partial_jK_{kl}\,\,.
\end{equation}
Using the fact that the term ${-\threeG{m}{j}{k}}K_{ml}$ in the covariant derivative $D_j K_{kl}$ 
vanishes once contracted with $\epsilon_{ijk}$, we arrive exactly at the operator 
of \eq{three_quadratic_parity_odd-1} up to irrelevant factors.

\section{Cubic graviton action from \texorpdfstring{$\vec{S_{\Lambda_3}}$}{S\_\{\textbackslash Lambda\_3\}}} 
\label{app:cubic_action_3DCS}

\noindent In this appendix we write down the expansion of 
\begin{equation}
\label{eq:app_cubic_action_3DCS-1}
\epsilon_{ijk}\,\Bigg(\frac{\threeG{l}{i}{m}\partial_j\threeG{m}{k}{l}}{2} 
+ \frac{\threeG{l}{i}{m}\threeG{m}{j}{n}\threeG{n}{k}{l}}{3}\Bigg) 
\end{equation} 
at cubic order in $\gamma_{ij}$, and how each term contributes to the amplitude in \eq{S_Lambda_3_in_in-A}. 

We have that, dropping total spatial derivatives \eq{app_cubic_action_3DCS-1} is the sum of $\smash{7}$ terms, i.e.~ 
\begin{equation}
\label{eq:app_cubic_action_3DCS-2}
\begin{split} 
\text{\eq{app_cubic_action_3DCS-1}} &= \frac{1}{4}\epsilon_{ijk}\gamma_{kn}\partial_l\gamma_{nm}\partial_j\partial_l\gamma_{im} 
+ \frac{1}{4}\epsilon_{ijk}\gamma_{ln}\partial_n\gamma_{im}\partial_j\partial_l\gamma_{km} 
- \frac{1}{4}\epsilon_{ijk}\partial_m\gamma_{nj}\partial_n\gamma_{lm}\partial_i\gamma_{lk} \\
&\;\;\;\; - \frac{1}{12}\epsilon_{ijk}\partial_m\gamma_{lj}\partial_n\gamma_{mi}\partial_l\gamma_{nk} 
+ \frac{1}{4}\epsilon_{ijk}\partial_n\gamma_{lm}\partial_k\gamma_{ln}\partial_j\gamma_{mi} 
+ \frac{1}{4}\epsilon_{ijk}\partial_m\gamma_{nj}\partial_m\gamma_{lk}\partial_l\gamma_{ni} \\ 
&\;\;\;\; + \frac{1}{4}\epsilon_{ijk}\partial_n\gamma_{mk}\partial_l\gamma_{mn}\partial_j\gamma_{li}\,\,. 
\end{split} 
\end{equation} 
Then, we can also see how these $\smash{7}$ terms contribute to the amplitude 
\begin{equation}
\label{eq:app_cubic_action_3DCS-3}
\epsilon^{s_1}_{il}({-\vec{k}_1}) 
\epsilon^{s_2}_{jm}({-\vec{k}_2}) 
\epsilon^{s_3}_{kn}({-\vec{k}_3}) 
T_{ijk}^{lmn}(\vec{k}_1,\vec{k}_2,\vec{k}_3) 
\end{equation} 
in \eq{S_Lambda_3_in_in-A}. More precisely, we list their contribution to $\smash{T_{ijk}^{lmn}(\vec{k}_1,\vec{k}_2,\vec{k}_3)}$, 
without writing down explicitly all six permutations that make it 
have the correct symmetry properties:\footnote{In each of the $\smash{7}$ terms of \eq{app_cubic_action_3DCS-2} 
we have chosen $\smash{(s_1,\vec{k}_1)}$ for the first field appearing in the cubic operator, 
$\smash{(s_2,\vec{k}_2)}$ for the second, and $\smash{(s_3,\vec{k}_3)}$ for the third.} 
\begin{enumerate}[leftmargin=*] 
\item ${-\dfrac{\iu}{4}}\epsilon_{iak}(\vec{k}_2\cdot\vec{k}_3)\delta_{mn}\delta_{lj}k^a_3$\,\,;
\item ${-\dfrac{\iu}{4}}\epsilon_{ajk}\delta_{mn}k^i_2k^a_3k^l_3$\,\,;
\item ${-\dfrac{\iu}{4}}\epsilon_{aik}\delta_{jn}k^m_1k^l_2k^a_3$\,\,;
\item ${\dfrac{\iu}{12}}\epsilon_{ijk}k^m_1k^n_2k^l_3$\,\,;
\item ${-\dfrac{\iu}{4}}\epsilon_{kab}\delta_{ij}\delta_{ln}k^m_1k^a_2k^b_3$\,\,;
\item ${\dfrac{\iu}{4}}\epsilon_{ijk}(\vec{k}_1\cdot\vec{k}_2)\delta_{ln}k^m_3$\,\,;
\item ${\dfrac{\iu}{4}}\epsilon_{aik}\delta_{lm}k^j_1k^n_2k^a_3$\,\,.
\end{enumerate}

Finally, once we write the amplitude as in \eqsII{S_Lambda_3_in_in-D}{S_Lambda_3_in_in-E}, i.e.~for the 
circularly-polarized tensors, we have that the fifth and seventh term in \eq{app_cubic_action_3DCS-2} 
contribute in the same way. Taking this into account, the first six terms then give exactly the six 
amplitudes $\smash{A^{(a)}_{S_{\Lambda_3}}(1^{s_1},2^{s_2},3^{s_3})}$ of \eqsI{S_Lambda_3_in_in-E}.

\section{Integrating by parts operators involving 
\texorpdfstring{$\vec{{{^{(3)}}\!R_{ij}}$}}{\{{\^{}}3\}R\_\{ij\}}
and \texorpdfstring{$\vec{\dot{K}_{ij}}$}{\textbackslash dot\{K\}\_\{ij\}}} 
\label{app:appendix-0}

\noindent In this appendix we review how to derive \eq{IBP_at_NNLO} and how to integrate by parts 
the parity-odd operator of \eq{NNLO-D-1}. As a warm-up let us derive the equation \cite{Gleyzes:2013ooa} 
\begin{equation}
\label{eq:five_operators-B}
\lambda(t){{^{(3)}}\!R_{\mu\nu}} K^{\mu\nu} = \frac{\lambda(t)}{2}{{^{(3)}}\!R}K 
+ \frac{\dot{\lambda}(t)}{2N}{{^{(3)}}\!R} + \text{boundary terms}\,\,. 
\end{equation} 
We rewrite $\smash{{{^{(3)}}\!R^{\mu\nu}}K_{\mu\nu}}$ as 
\begin{equation}
\label{eq:appendix-0-A}
\begin{split}
\smash{{{^{(3)}}\!R^{\mu\nu}}K_{\mu\nu}} &= \frac{1}{2}{{^{(3)}}\!R^{\mu\nu}}\mathcal{L}_{\vec{n}}h_{\mu\nu} \\
&= \frac{1}{2}\mathcal{L}_{\vec{n}}{{^{(3)}}\!R} - h_{\mu\nu}\mathcal{L}_{\vec{n}}{{^{(3)}}\!R^{\mu\nu}} \\
&= \frac{1}{2}\nabla_\rho\big({{^{(3)}}\!R} n^\rho\big) - \frac{1}{2}{{^{(3)}}\!R}K 
- h_{\mu\nu}\mathcal{L}_{\vec{n}}{{^{(3)}}\!R^{\mu\nu}}\,\,.
\end{split}
\end{equation}
It is then only a matter of computing the Lie derivative of ${{^{(3)}}\!R^{\mu\nu}}$ along $n^\rho$. 
We can rewrite it in terms of the change of ${{^{(3)}}\!R^{\mu\nu}}$ with respect to a variation $\Delta h_{\rho\sigma}$ 
evaluated at $\Delta h_{\rho\sigma} = \mathcal{L}_{\vec{n}}h_{\rho\sigma} = 2K_{\rho\sigma}$. Using the relation 
$\smash{\Delta h^{\mu\nu} = -h^{\mu a} h^{\nu b}\Delta h_{\mu\nu}}$, together with 
\begin{equation}
\label{eq:appendix-0-B}
\Delta\!{{^{(3)}}\!R_{\mu\nu}} = -\frac{1}{2}D_\rho D^\rho\Delta h_{\mu\nu} 
+ \frac{1}{2}h^\rho_{\hp{\rho}\mu}D^\sigma D_\nu\Delta h_{\rho\sigma} 
+ \frac{1}{2}h^\rho_{\hp{\rho}\nu}D^\sigma D_\mu\Delta h_{\rho\sigma}\,\,,
\end{equation}
we get 
\begin{equation}
\label{eq:appendix-0-C}
\begin{split}
\smash{{{^{(3)}}\!R^{\mu\nu}}K_{\mu\nu}} &= \frac{1}{2}{{^{(3)}}\!R}K - \frac{1}{2}\nabla_\rho\big({{^{(3)}}\!R} n^\rho\big) 
+ h^{\mu\nu}\Delta\!{{^{(3)}}\!R_{\mu\nu}}|_{\Delta h_{\rho\sigma} = 2K_{\rho\sigma}}\,\,,
\end{split}
\end{equation}
where the last term is a total spatial divergence. It can be neglected under 
$\smash{\int\dif^4x\,\sqrt{-g}}$ up to terms involving scalar modes. 

We can manipulate the operator $\smash{{{^{(3)}}\!R_{\mu\nu}}n^\rho\nabla_\rho K^{\mu\nu}}$ in a similar way 
(notice that $\smash{n^\rho\nabla_\rho K^{\mu\nu}}$ is equal to $\smash{n^\rho\nabla_\rho \delta\!K^{\mu\nu}}$ 
up to terms suppressed by $\varepsilon$ and terms involving $\smash{A^\mu}$). First, we write 
\begin{equation}
\label{eq:appendix-0-D}
{{^{(3)}}\!R_{\mu\nu}}n^\rho\nabla_\rho K^{\mu\nu} = {{^{(3)}}\!R_{\mu\nu}}\mathcal{L}_{\vec{n}} K^{\mu\nu} 
+ 2{{^{(3)}}\!R_{\mu\nu}} K^{\mu\rho}K_{\rho}^{\hp{\rho}\nu}\,\,. 
\end{equation}
Expanding $K_{\mu\nu} = Hh_{\mu\nu} + \delta\!K_{\mu\nu}$ we see that the second term on the right-hand side 
generates lower-derivative operators and modifies the coefficient of the cubic operator of \eq{NNLO-B-2}. 
What about the first term? We perform the same integrations by parts that led to \eq{appendix-0-C}. 
The new piece we have is $\smash{{-K^{\mu\nu}}\mathcal{L}_{\vec{n}}{{^{(3)}}\!R_{\mu\nu}}}$. Using 
\eq{appendix-0-B} with $\Delta h_{\rho\sigma} = 2K_{\rho\sigma}$, it is equal to 
\begin{equation}
\label{eq:appendix-0-E}
{-2K^{\rho\sigma}D_\mu D_\sigma K^\mu_{\hp{\mu}\rho}} + K^{\rho\sigma}D^\mu D_\mu K_{\rho\sigma}\,\,. 
\end{equation} 
We recognize exactly the second operator of \eq{NNLO-A-2}, and we can rewrite the first term as 
\begin{equation}
\label{eq:appendix-0-F}
\begin{split}
K^{\rho\sigma}D_\mu D_\sigma K^\mu_{\hp{\mu}\rho} &= K^{\rho\sigma}D_\sigma D_\mu K^\mu_{\hp{\mu}\rho} 
+ K^{\rho\sigma}[D_\mu, D_\sigma] K^\mu_{\hp{\mu}\rho} \\
&= D_\sigma(K^{\rho\sigma} D_\mu K^\mu_{\hp{\mu}\rho}) - D_\sigma K^{\rho\sigma} D_\mu K^\mu_{\hp{\mu}\rho} 
+ {{^{(3)}}\!R_{\rho\sigma}}K^{\lambda\sigma}K^\rho_{\hp{\rho}\lambda} 
- {{^{(3)}}\!R^\rho_{\hp{\rho}\lambda\mu\sigma}}K^{\lambda\sigma}K^\mu_{\hp{\mu}\rho}\,\,.
\end{split}
\end{equation}
The last two terms modify the coefficients of the two cubic operators of \eqsI{NNLO-B} 
and of lower-derivative operators, while the second term starts at fourth order in perturbations if we consider only tensor modes. 

We conclude this appendix by showing that the operator ${\vec{e}^{\mu\nu\rho\sigma}n_\mu} 
D_\nu\delta\!K_{\rho\lambda}n^\delta\nabla_\delta\delta\!K^\lambda_{\hp{\lambda}\sigma}/N$ 
can also be removed. Again, we drop scalar modes throughout. First, we use the relation 
\begin{equation} 
\label{eq:appendix-0-G}
n^\delta\nabla_\delta K^\lambda_{\hp{\lambda}\sigma} = {\cal L}_{\vec{n}}K^\lambda_{\hp{\lambda}\sigma}\,\,.
\end{equation}
With this relation we rewrite the operator as 
\begin{equation}
\label{eq:appendix-0-H}
\begin{split}
\frac{\vec{e}^{\mu\nu\rho\sigma}n_\mu}{N}\,D_\nu\delta\!K_{\rho\lambda}n^\delta\nabla_\delta
\delta\!K^\lambda_{\hp{\lambda}\sigma} &= \frac{\vec{e}^{\mu\nu\rho\sigma}n_\mu}{N}\,D_\nu K_{\rho\lambda} 
n^\delta\nabla_\delta K^\lambda_{\hp{\lambda}\sigma} \\
&= {\cal L}_{\vec{n}}\bigg(\frac{\vec{e}^{\mu\nu\rho\sigma}n_\mu}{N}\, 
D_\nu K_{\rho\lambda} K^\lambda_{\hp{\lambda}\sigma}\bigg) - K^\lambda_{\hp{\lambda}\sigma}\, 
{\cal L}_{\vec{n}}\bigg(\frac{\vec{e}^{\mu\nu\rho\sigma}n_\mu}{N}\,D_\nu K_{\rho\lambda}\bigg)\,\,. 
\end{split}
\end{equation}
In the first term on the right-hand side we recognize the operator of $\smash{S_{\Lambda_2}}$ 
after integrating by parts. What about the second term? 
Using the relations $\smash{{\cal L}_{\vec{n}} n_\mu = A_\mu}$ and $\smash{{\cal L}_{\vec{n}}\vec{e}^{\mu\nu\rho\sigma} 
= {-K} \vec{e}^{\mu\nu\rho\sigma}}$ we see that we only care about the Lie derivative of 
$\smash{D_\nu K_{\rho\lambda}}$. When the Lie derivative acts on $K_{\rho\lambda}$ we get ${-{\vec{e}^{\mu\nu\rho\sigma}n_\mu}} 
D_\nu K_{\rho\lambda}n^\delta\nabla_\delta K^\lambda_{\hp{\lambda}\sigma}/N$ (notice the minus sign) plus 
terms starting cubic in perturbations that are reabsorbed by the operator of \eq{NNLO-E}. Finally, using the relation 
\begin{equation}
\label{eq:appendix-0-I}
\Delta\!{^{(3)}}\Gamma^{\alpha}_{\hp{\rho}\mu\nu}|_{\Delta h_{\rho\sigma} = 2K_{\rho\sigma}} = 
{-D^\alpha}K_{\mu\nu} + D_\mu K_\nu^{\hp{\nu}\alpha} + D_\nu K_\mu^{\hp{\mu}\alpha} 
\end{equation}
for the (projected) variation of the Christoffel symbols in $D_\nu$, we see that 
when ${\cal L}_{\vec{n}}$ acts on $D_\nu$ we again get terms that ``renormalize'' the coefficient of the operator of \eq{NNLO-E}.

\section{Stueckelberg trick for \texorpdfstring{$\vec{\epsilon_{ijk}D_iK_{jl}K^l_{\;k}}$}
{\textbackslash epsilon\_\{ijk\} D\_iK\_\{jl\} K{\^{}}l\_k}}
\label{app:appendix-C}

\noindent In this appendix we show how to perform the Stueckelberg trick for 
the operator in $\smash{S_{\Lambda_2}}$. Let us define $\smash{{\cal O}_{\Lambda_2} = 2{\cal O}_{1,2}}$, 
cf.~\eqsII{summary_redundancies-B-2}{three_quadratic_parity_odd-1}, as 
\begin{equation}
\label{eq:eq_appendix-C-1-a}
{\cal O}_{\Lambda_2} = \frac{\vec{e}^{\mu\nu\rho\sigma}n_\mu}{N}\,D_\nu\delta\!K_{\rho\lambda} 
\delta\!K^\lambda_{\hp{\lambda}\sigma}\,\,. 
\end{equation}
Since $\Lambda_2$ is constant in time we do not need to consider it when we introduce the $\pi$ field. 

In order to perform the Stueckelberg trick, we first rewrite $\smash{{\cal O}_{\Lambda_2}}$ in a way that 
involves as few tensors that are not invariant under time diffeomorphisms as possible. 
First, notice that the definition of \eq{eq_appendix-C-1-a} is equivalent to 
\begin{equation}
\label{eq:eq_appendix-C-2}
{\cal O}_{\Lambda_2} = \frac{\vec{e}^{\mu\nu\rho\sigma}n_\mu}{N}\,D_\nu K_{\rho\lambda}\, K^\lambda_{\hp{\lambda}\sigma} 
= \frac{\vec{e}^{\mu\nu\rho\sigma}n_\mu}{N}\,D_\nu K_{\rho\lambda}\nabla^\lambda n_\sigma\,\,. 
\end{equation}
Then, we use the Codazzi-Mainardi relation 
\begin{equation}
\label{eq:eq_appendix-C-3}
h^\nu_{\hp{\nu}a}h^b_{\hp{b}\lambda}n^ch^d_{\hp{d}\rho}R^a_{\hp{a}dbc} = D^\nu K_{\rho\lambda} - D_\rho K^\nu_{\hp{\nu}\lambda} 
\end{equation}
together with the antisymmetry of $\vec{e}^{\mu\nu\rho\sigma}$ to rewrite ${\cal O}_{\Lambda_2}$ as 
\begin{equation}
\label{eq:eq_appendix-C-4-a}
{\cal O}_{\Lambda_2} = \frac{1}{2}\frac{\vec{e}^{\mu\nu\rho\sigma}n_\mu}{N}\,h^b_{\hp{b}\lambda} 
n^c R_{\nu\rho bc} \nabla^\lambda n_\sigma\,\,.
\end{equation}
Finally, employing the antisymmetry of $R_{\nu\rho bc}$ in $bc$ we arrive at 
\begin{equation}
\label{eq:eq_appendix-C-4-b}
{\cal O}_{\Lambda_2} = \frac{1}{2}\vec{e}^{\alpha\beta\rho\sigma}R_{\rho\sigma\mu\nu}n_\alpha 
n^\nu\nabla^\mu\bigg(\frac{n_\beta}{N}\bigg)\,\,.
\end{equation}
The advantage of \eq{eq_appendix-C-4-b} is that we only need to know how $n_\mu$ and $N$ transform under $t\to t+\pi$. 
Using that $n_\mu = -N\partial_\mu t$ and that $g^{00} = g^{\mu\nu}\partial_\mu t\partial_\nu t = {-1/N^2}$ we have 
\begin{subequations}
\label{eq:eq_appendix-C-5}
\begin{align}
N &\to\tilde{f}\cdot N\,\,, \label{eq:eq_appendix-C-5-1} \\
n_\mu &\to\tilde{f}\cdot(n_\mu - N\nabla_\mu\pi)\equiv\tilde{f}\cdot p_\mu\,\,, \label{eq:eq_appendix-C-5-2} \\
\tilde{f} &= \frac{1}{\sqrt{1+2Nn^\mu\nabla_\mu\pi-N^2g^{\mu\nu}\nabla_\mu\pi\nabla_\nu\pi}} 
\label{eq:eq_appendix-C-5-3}
\end{align}
\end{subequations}
and 
\begin{equation}
\label{eq:eq_appendix-C-6}
{\cal O}_{\Lambda_2}\to\frac{\tilde{f}^2}{2}\vec{e}^{\alpha\beta\rho\sigma}R_{\rho\sigma\mu\nu}p_\alpha 
p^\nu\nabla^\mu\bigg(\frac{p_\beta}{N}\bigg)\,\,.
\end{equation}

First, let us confirm that $\smash{{\cal O}_{\Lambda_2}}$ does not couple $\pi$ 
to the constraints. Using ${{^{(3)}}\!R_{\nu[\beta\rho\sigma]}} = 0$, 
the only second-order term is of the form 
$\smash{\vec{e}^{\alpha\beta\rho\sigma}n_\alpha D_\rho\delta\!K_{\sigma\mu}D^\mu D_\beta\pi\propto 
\epsilon_{ijk}D_j\delta\!K_{kl}D^lD_i\pi}$. Since at this order 
$\delta\!K_{kl} = {-H}\delta\!N h_{kl} - D_{(k}N_{l)} = {-H}\delta\!N h_{kl} - D_{k}N_{l}$ (thanks to 
the fact that $D_i$ is torsionless and that $N_i = D_i N_{\rm L}$), we remain with 
\begin{equation}
\label{eq:mixing}
\epsilon_{ijk}D_j\delta\!K_{kl}D^lD_i\pi\supset{-\epsilon_{ijk}}D_jD_kN_lD^lD_i\pi\,\,.
\end{equation}
At this order in perturbations we can exchange $D_jD_k$ with $D_kD_j$, hence this term vanishes. 

Since there is no additional mixing of $\pi$ with the constraints, the correction to Maldacena's solution for 
$\delta\!N$ and $N_{\rm L}$ enters only at order $\varepsilon H/\Lambda$. 
According to the discussion in Section~\ref{subsec:estimates}, at leading order in slow roll 
we can then put $N=1$, $N^i=0$ in $\pi$ gauge when discussing non-Gaussianities involving scalar modes. 
\eqsIII{eq_appendix-C-5-2}{eq_appendix-C-5-3}{eq_appendix-C-6} then become 
\begin{subequations}
\label{eq:eq_appendix-C-7}
\begin{align}
p_\mu &= \tilde{f}\cdot(n_\mu-\nabla_\mu\pi)\,\,, \label{eq:eq_appendix-C-7-1} \\
\tilde{f} &= \frac{1}{\sqrt{1+2\dot{\pi}-\dot{\pi}^2 + h^{\mu\nu}D_\mu\pi D_\nu\pi}} 
\label{eq:eq_appendix-C-7-2}
\end{align}
\end{subequations}
and 
\begin{equation}
\label{eq:eq_appendix-C-8}
{\cal O}_{\Lambda_2}\to\frac{\tilde{f}^2}{2}\vec{e}^{\alpha\beta\rho\sigma}R_{\rho\sigma\mu\nu}p_\alpha p^\nu\nabla^\mu{p_\beta}\,\,. 
\end{equation} 
In terms of metric fluctuations, up to cubic order and assuming a constant Hubble rate 
we then find that ${\cal O}_{\Lambda_2}$ is equal to $\tilde{f}^2/2$ times the sum of three terms: 
\begin{enumerate}[leftmargin=*]
\item from the Gauss relation $h^c_{\hp{c}\rho}h^d_{\hp{d}\sigma}h^a_{\hp{a}\mu}h^b_{\hp{b}\nu}R_{cdab} 
= {{^{(3)}}\!R_{\rho\sigma\mu\nu}} + 2K_{\rho[\mu}K_{\nu]\sigma}$ we get 
\begin{equation}
\label{eq:eq_appendix-C-9}
\begin{split}
&\frac{a^{-2}}{2}\vec{e}^{0ijk}{{^{(3)}}\!R_{jklm}}\partial_m\pi(\dot{\gamma}_{li} 
- 2a^{-2}\partial_l\partial_i\pi) - H\vec{e}^{0ijk}\dot{\gamma}_{jl}\partial_k\pi\partial_l\partial_i\pi\,\,;
\end{split}
\end{equation}
\item from the Codazzi-Mainardi relation we have 
\begin{equation}
\label{eq:eq_appendix-C-10}
\begin{split}
&2(1+\dot{\pi})^2\vec{e}^{0ijk}D_k\delta\!K_{jm}\big((1+\dot{\pi})\delta\!K^m_{\hp{m}i} - D^mD_i\pi\big) \\
&+ 2\vec{e}^{0ijk}\partial_k\dot{\gamma}_{jl}\partial_{(l}\pi\partial_{i)}\dot{\pi} 
- H\vec{e}^{0ijk}\partial_k\dot{\gamma}_{jl}\partial_{l}\pi\partial_{i}{\pi}\,\,;
\end{split}
\end{equation}
\item from the Ricci equation $h^c_{\hp{c}\rho}n^dh^a_{\hp{a}\mu}n^bR_{cdab} = {-K_{\rho\delta}K^\delta_{\hp{\delta}\mu}} 
+ D_\mu A_\rho -n^\delta\nabla_\delta K_{\rho\mu} + 2n_{(\rho} K_{\mu)\delta} A^\delta + A_\rho A_\mu$ we get 
\begin{equation}
\label{eq:eq_appendix-C-11}
\vec{e}^{0ijk}\bigg(a^2H\dot{\gamma}_{kl} + \frac{a^2}{2}\ddot{\gamma}_{kl}\bigg)\partial_i\pi\, 
(2a^{-2}\partial_l\partial_j\pi - \dot{\gamma}_{lj})\,\,.
\end{equation}
\end{enumerate}
Here $\vec{e}^{0ijk} = -\epsilon_{ijk}/\sqrt{-g}$ and $\tilde{f} = 1-\dot{\pi} + 2\dot{\pi}^2 - a^{-2}\partial_i\pi\partial_i\pi/2$. 
In \eq{eq_appendix-C-10} we have that 
\begin{subequations}
\label{eq:eq_appendix-C-12}
\begin{align}
D_k\delta\!K_{jm} &= \partial_k\delta\!K_{jm} - \threeG{l}{k}{j}\delta\!K_{lm} - \threeG{l}{k}{m}\delta\!K_{jl}\,\,, 
\label{eq:eq_appendix-C-12-1} \\
D^mD_i\pi &= a^{-2}(\eu^{-\gamma})_{mn}\big(\partial_n\partial_i\pi - \threeG{l}{n}{i}\partial_l\pi\big)\,\,, 
\label{eq:eq_appendix-C-12-2} \\
{{^{(3)}\!}R}_{jklm} &= a^2\big(\partial_l\threeG{j}{m}{k}-\partial_m\threeG{j}{l}{k}\big)\,\,, \label{eq:eq_appendix-C-12-3} \\
\delta\!K_{jm} &= \frac{a^2}{2}\frac{\dif (\eu^\gamma)_{jm}}{\dif t}\,\,, \label{eq:eq_appendix-C-12-4} \\
\delta\!K^m_{\hp{m}i} &= \frac{1}{2}(\eu^{-\gamma})_{mn}\frac{\dif (\eu^\gamma)_{ni}}{\dif t}\,\,, \label{eq:eq_appendix-C-12-5} \\
\threeG{k}{i}{j} &= \frac{{-\partial_k\gamma_{ij} + \partial_i\gamma_{jk} + \partial_j\gamma_{ki}}}{2}\,\,. \label{eq:eq_appendix-C-12-6}
\end{align}
\end{subequations}
From these formulas we see that the operator ${\cal O}_{\Lambda_2}$ contains direct interactions of the graviton with $\pi$, 
i.e.~there are $\pi\gamma\gamma$ and $\gamma\pi\pi$ vertices even in absence of scalar metric fluctuations.



\clearpage

\bibliographystyle{utphys}
\bibliography{refs}


\end{document}